\documentclass[journal=gmj]{CUP-JNL-DTM}%

\usepackage{graphicx}
\usepackage{multicol,multirow}
\usepackage{amsmath,amssymb,amsfonts}
\usepackage{mathrsfs}
\usepackage{amsthm}
\usepackage{rotating}
\usepackage{appendix}
\usepackage{ifpdf}
\usepackage[T1]{fontenc}
\usepackage{newtxtext}
\usepackage{newtxmath}
\usepackage{textcomp}
\usepackage{xcolor}
\usepackage{lipsum}
\usepackage[colorlinks,allcolors=blue]{hyperref}

\DeclareMathOperator{\pt}{\partial}

\DeclareMathOperator*{\argmin}{argmin}

\usepackage{algorithm}
\usepackage{array}
\usepackage{pdflscape}
\usepackage[table]{xcolor}

\newcolumntype{L}[1]{>{\raggedright\arraybackslash}p{#1}}
\newcommand{\resultcell}[1]{\(\begin{cases} #1 \end{cases}\)}
\newcommand{\resultrow}[5]{#1 & #2 & #3 & #4 & #5 \\[5pt]}
\newcommand{\smallresultrow}[4]{#1 & #2 & #3 & #4 \\[5pt]}

\newcommand{\customdashrule}[1]{%
  \makebox[#1][l]{%
    {\color{black!35}%
    \leaders\hbox{\rule[0.55ex]{3pt}{0.35pt}\hskip 2.5pt}\hfill}%
  }%
}
\newcommand{\lightdashline}{%
  \noalign{\vskip -2.5pt}%
  \multicolumn{5}{@{}l@{}}{\customdashrule{1\linewidth}}\\[-2pt]%
  \noalign{\vskip 1pt}%
}

\theoremstyle{definition}

\numberwithin{equation}{section}

\jname{}
\articletype{}
\jyear{2026}

\begin{document}

\begin{Frontmatter}

\title[Article Title]{Learning Effective Soliton Dynamics from Scattering Data}

\author[1,2]{Seth Minor}
\author[1]{Vanja Dukic}
\author[1]{David M.~Bortz}

\authormark{Minor \textit{et al}.}

\address[1]{\orgdiv{Department of Applied Mathematics}, \orgname{University of Colorado}, \orgaddress{\city{Boulder}, \state{CO},  \country{USA}}}

\address[2]{\email{seth.minor@colorado.edu}}

\authormark{Minor et al.}



\abstract{
The \textit{inverse scattering transform} (IST) provides the standard theoretical framework for deriving soliton dynamics. Traditionally, such derivations have been of an analytical, rather than data-driven, nature. In this paper, we combine the conceptual framework of the IST with weak-form system identification methods to discover effective soliton dynamics directly from observed scattering data, without assuming prior knowledge of the scattering equations. Our method avoids parameterizing solitary waves via ad hoc curve-fitting by working in the scattering domain, yielding interpretable low-dimensional models that remain valid in perturbed and near-integrable regimes. We demonstrate the performance of the proposed approach on synthetic and experimental data governed by shallow-water equations of Korteweg--de Vries-type and recover models that are consistent with canonical IST theory.
}

\end{Frontmatter}


\localtableofcontents

\clearpage

\section{Introduction}
A central mathematical problem in the study of nonlinear dispersive waves is the construction of effective, low-dimensional models for coherent solitary wave structures (\textit{solitons}). In such settings, the \textit{inverse scattering transform} (IST) of Ablowitz, Kaup, Newell, and Segur \cite{Ablowitz.etal1974StudApplMath} has enjoyed a rich and successful history, and is now the standard theoretical framework for deriving reduced-order evolution equations for soliton dynamics. Although these derivations are traditionally of an analytical -- rather than data-driven -- nature, recent work has employed the IST formalism as a tool for experimental data analysis, using the technique to analyze soliton content from empirical measurements \cite{Feng.etal2023FrontPhys, Lee.etal2024PLoSONE, Tikan.etal2022SciRep}. Moreover, recent approaches using alternative parameterization techniques have demonstrated that the learning of reduced-order, interpretable equations of motion for solitons is tenable in a data-driven setting \cite{Chen.etal2025ChaosInterdiscipJNonlinearSci, Yang.etal2024FrontPhotonics, Yang.etal2025}. Despite the success of this recent work, however, little effort has been devoted to developing a data-driven modeling approach based on the IST itself, most likely due to the fact that the framework is fundamentally problem-specific.

\smallskip

In this paper, we address the question of whether effective soliton dynamics can be inferred directly from observed scattering data (as opposed to being derived or approximated analytically). We show that the conceptual framework of the IST can be combined with modern weak-form equation learning methods to identify interpretable models for the scattering dynamics, without assuming prior knowledge about the form of the equations. Our proposed method avoids parameterizing solitary waves via ad hoc curve-fitting by working in natural spectral domains, yielding interpretable low-dimensional models that remain valid in perturbed, near-integrable regimes. In this paper, we specifically focus on examples related to the canonical Korteweg--de Vries equation; however, we do briefly comment on the potential for extension to other integrable equations in Section~\ref{sec:discussion}.

\smallskip

The paper is organized as follows. In Section~\ref{sec:preliminaries}, we relate to previous work (\S\ref{subsec:previous_work}), establish some notational conventions and background assumptions (\S\ref{subsec:notation_and_assumptions}), identify the specific class of models being considered (\S\ref{subsec:equations_considered}), and review background material on the IST (\S\ref{subsec:IST}) as well as the paradigm of weak form modeling (\S\ref{subsec:wsindy}). Later, in Section~\ref{sec:methods}, we outline our approach (\S\ref{subsec:outline}) and numerical implementation (\S\ref{subsec:numerical_implementation} and \S\ref{subsec:candidate_library}). In Section~\ref{sec:results}, we discuss validation metrics (\S\ref{subsec:validation}) before presenting some preliminary numerical results obtained using simulated data (\S\ref{subsec:synthetic_data}) and experimental water-wave data (\S\ref{subsec:empirical_data}), pausing briefly to comment on structural identifiability concerns (\S\ref{subsec:identifiability}). Finally, we conclude with a summary and discussion in Section~\ref{sec:discussion}. Supplemental information is given in the Appendix.

\section{Background Material}\label{sec:preliminaries}

\subsection{Relation to Previous Work}\label{subsec:previous_work}
It is important to note that there are many ways of parameterizing solitary waves, and different choices can lead to distinct effective models \cite{Guo.etal2022PhysRevResearch, Zhang.etal2025ComputerMethodsinAppliedMechanicsandEngineering}. In the recent literature, Chen, Yang, Zhu, and Kevrekidis have explicitly fit parameterized waveforms (centers, widths, amplitudes, phases) to simulated data \cite{Yang.etal2025}, and have also considered parameterizing the statistical moments of physically relevant quantities \cite{Chen.etal2025ChaosInterdiscipJNonlinearSci, Yang.etal2024FrontPhotonics}, before then applying equation learning approaches to discover effective dynamic models governing the corresponding parameters. Here, we instead propose to use \textit{scattering data} (formally defined in \S\ref{subsec:IST}) directly as a parameterization of the soliton field. This choice has at least two advantages: \begin{itemize}
    \item it is natural in the context of nonlinear wave equations, inheriting a conceptual framework and set of known analytical results from existing \textit{direct}/\textit{inverse scattering transform} (DST/IST) theory;

    \smallskip
    
    \item it avoids fitting ad hoc, problem-specific waveforms to the data.
\end{itemize} Unfortunately, there is no free lunch regarding the latter point -- a significant caveat of using scattering data is that, while the IST framework of \cite{Ablowitz.etal1974StudApplMath} does supply a fairly unified language across many integrable equations of interest, the definitions of both DST and IST are inherently problem-specific.

\newpage

\subsection{Notation and Assumptions}\label{subsec:notation_and_assumptions}
Herein, we will let $u = u(x,t)$ denote a scalar-valued field which represents a solution to a perturbed, dispersive partial differential equation (PDE) of the form \begin{align}\label{eq:dispersive_PDE}
    u_t + N[u] = \epsilon F[u],
    \quad \text{with} \quad
    \begin{cases}
        u(x,0) = u_0(x),
        \\
        (x,t) \in (-\infty, \infty) \times [0,T],
    \end{cases}
\end{align} where $N$ and $F$ are (potentially nonlinear) differential operators and $0 \leq \epsilon \ll 1$ is a small parameter representing a perturbation to the system. Of particular interest are \textit{integrable systems},\footnote{For our purposes, it will suffice to define an `integrable system' as a system of partial differential equations that admits a Lax pair representation.} which for $\epsilon = 0$ admit a \textit{Lax pair} -- linear operators $L=L(u)$ and $M=M(u)$ which depend on $u$ and satisfy \begin{align}\label{eq:lax_eqn}
    \big(L_t(u) + [L, M](u)\big)\phi = \big(u_t + N[u]\big) \cdot \phi = 0,
\end{align} for any test function $\phi = \phi(x,t)$ of appropriate smoothness, where $[\cdot\,,\cdot]$ is the commutator operator and $L_t := [\pt_t, L]$ denotes the operator-valued time derivative of $L$. This identity arises as a consistency condition for an associated overdetermined eigenvalue problem (EVP) given by \begin{align}\label{eq:lax_evp}
    \begin{cases}
        L(u)\phi = \lambda\phi,
        \\
        \phi_t = M(u)\phi.
    \end{cases}
\end{align} To keep the exposition somewhat concise and self-contained, we have deferred a more detailed review of Lax pairs to Appendix~\S\ref{appendixB:lax_pairs}. Throughout the paper, the following two assumptions are in play: (1) the field decays asymptotically (i.e., $|u| \rightarrow 0$ as $x \rightarrow \pm\infty$),\footnote{From an analytical perspective, the rate of decay must be sufficiently fast -- a standard assumption is that $u \in L^1$ and $(1+|x|) \cdot u_0 \in L^1$.} and (2) the underlying PDE and Lax pair are available in closed form. Regarding (2), we note that, e.g., the WSINDy \cite{Messenger.etal2024SciRep} and SILO \cite{Adriazola.etal2026SIAMJApplDynSyst} algorithms are respectively capable of identifying the $N$ and $L,M$ operators from nonlinear wave equation data, indicating that an end-to-end data-driven extension of the current work is also tenable.

\vspace{-2.5 mm}

\subsection{Equations Considered}\label{subsec:equations_considered}
To illustrate our proposed method on a canonical and well-understood testbed, we focus on examples related to the \textit{Korteweg-de Vries} (KdV) \textit{equation}, a prototypical model of shallow water waves given by \begin{align}\label{eq:kdv_eqn}
    u_t + \alpha uu_x + \beta u_{xxx} = 0,
    \quad \text{where} \quad
    \begin{cases}
        L(u) = -\big(\!\pt^2_{\!x} + \, \frac{\alpha}{6\beta}u\big),
        \\
        M(u) = -\big(4\beta\!\pt^3_{\!x} + \, \alpha u\!\pt_{\!x} + \, \frac{\alpha}{2}[\pt_{\!x}, u]\big).
    \end{cases}
\end{align} We also consider nontrivial perturbations of the KdV equation, leading to models of the form \begin{align}\label{eq:perturbed_kdv_eqn}
    u_t + \alpha uu_x + \beta u_{xxx} = \epsilon F[u].
\end{align} In particular, for a fifth-order dispersive forcing term $F[u] := \pt^5_{\!x} u$, the perturbed KdV model given above collapses to the \textit{Kawahara equation}, \begin{align}\label{eq:kawahara_eqn}
    u_t + \alpha uu_x + \beta u_{xxx} = \epsilon u_{xxxxx}.
\end{align} Our focus on this specific class of governing equations is partly motivated by the recent success of Heinrich et al. \cite{Heinrich.etal2025} in applying data-driven equation-learning methods to experimental wave-tank data, in which a Kawahara-type model was identified.\footnote{In \cite{Heinrich.etal2025}, the identified model was equivalent to eq.~(\ref{eq:kawahara_eqn}) with coefficients $\alpha \approx 1.36$, $\beta \approx 0.62$, and $\epsilon \approx 0.06$; see \S\ref{subsec:empirical_data} below.} However, a near-identical pipeline extends to other integrable equations that admit a Lax pair representation; we return to this point in
Section~\ref{sec:discussion}.

\newpage

\subsection{The Inverse Scattering Transform}\label{subsec:IST}
In standard acoustic scattering problems, one is often concerned with identifying the form of a spatially varying parameter $v(x)$ associated with a linear hyperbolic PDE such as the classical wave equation, \begin{align*}
    v(x)^{-2} u_{tt} - u_{xx} = 0,
\end{align*} in some inhomogeneous region $x \in \Omega$ where $v(x) \rightarrow v_0$ collapses to a background value as $x \rightarrow \infty$. Adopting a temporally-modulated ansatz of the form $u(x,t) = \phi(x) e^{-i\omega{t}}$ and heuristically introducing a \textit{scattering potential} $V(x) := k^2(1 - \eta(x)^2)$, where $\eta(x) := v_0/v(x)$ is the \textit{index of refraction} and $k := \omega/v_0$ is an associated wavenumber, one obtains a Schr\"odinger-type EVP (cf. eq.~(\ref{eq:lax_evp})) given by \begin{align}\label{eq:standard_scattering_EVP}
    L\phi = k^2\phi,
    \quad \text{with} \quad
    L = -\pt^2_{\!x} \, + \, V(x).
\end{align} For fixed $V(x)$, the self-adjoint operator $L$ has a point spectrum defined by a finite set of eigenvalues $\lambda_i = k^2_i < 0$ for $i = 1, \dots, n$, and a continuous spectrum which consists of all $\lambda = k^2 \in [0,\infty)$. Traditionally, the solutions to eq.~(\ref{eq:standard_scattering_EVP}) are decomposed linearly as $\phi = \phi_{\text{in}} + \phi_{\text{out}}$, where $\phi_{\text{in}}(x;k) = e^{\boldsymbol{i}kx}$ represents an incoming wave and $\phi_{\text{out}}(x;k_i) \sim c_ie^{\boldsymbol{i}k_ix}$ or $\phi_{\text{out}}(x;k) \sim R(k)e^{\boldsymbol{i}kx}$ represent bound or scattered waves corresponding to the point spectrum or the continuous spectrum, respectively.

\smallskip

It is often convenient to write the asymptotic weighting coefficients $c_i$ and $R(k)$, respectively referred to as the \textit{norming constants} and \textit{reflection coefficient}, in terms of the positive \textit{Jost solution}, $\phi_{+}$, of the Schr\"odinger EVP in eq.~(\ref{eq:standard_scattering_EVP}). For $k \neq 0$, this eigenfunction can be uniquely defined as the solution of the EVP subject to the boundary condition $\phi_{+}(x;k) = \rho(x;k)e^{ikx} \sim e^{\boldsymbol{i}kx}$ as $x \rightarrow \infty$, or equivalently: \begin{align}\label{eq:jost_soln}
    L\rho = 2\boldsymbol{i}k_i\rho',
    \quad \text{subject to} \quad
    \begin{cases}
        \rho(\infty) = 1,
        \\
        \rho'(\infty) = 0.
    \end{cases}
\end{align} Following the Marchenko convention and defining $\kappa := -\boldsymbol{i}k$ with $\lambda_i = -\kappa^2_i < 0$, one typically expresses \begin{align}\label{eq:norming_constant}
    c(\kappa_i) := \left[ \int_{-\infty}^{\infty} \phi_{+}(x,t;\boldsymbol{i}\kappa_i)^2 \, dx \right]^{-1}\!
    \quad \text{and} \ \quad
    R(k) := \frac{b(k)}{a(k)},
\end{align} given that $\phi_+(x;k) \sim a(k)e^{\boldsymbol{i}kx} + b(k)e^{-\boldsymbol{i}kx}$ as $x \rightarrow -\infty$. We refer to the set of $n$ eigenvalues $\lambda_i := -\kappa^2_i$ in the point spectrum of $L$, together with the corresponding constants $c_i := c(\kappa_i)$ and reflection coefficient $R(k)$ for each $\lambda = k^2$ in the continuous spectrum, as the \textit{scattering data}, $\sigma$, corresponding to the EVP: \begin{align}\label{eq:scattering_data_def}
    \sigma(L) := \big\{\big(\kappa_i, c(\kappa_i)\big) \, : \, i=1, \dots,n\big\} \, \cup \, \big\{ \big(k, R(k)\big) \, : \, k \in \mathbb{R} \big\}.
\end{align} In a \textit{forward scattering problem}, one aims to compute the scattering data for a known potential $V(x)$, while in an \textit{inverse scattering problem}, one instead aims to recover $V(x)$ from observed scattering data.

\smallskip

At a high level, the IST formalism of \cite{Ablowitz.etal1974StudApplMath} turns the standard scattering paradigm described above on its head by associating a nonlinear PDE of the form of eq.~(\ref{eq:dispersive_PDE}) with an equivalent linear EVP of the form of eq.~(\ref{eq:lax_evp}), by way of a Lax pair representation $L_t + [L,M] = u_t + N[u]$. For the shallow-wave models introduced in \S\ref{subsec:equations_considered}, this can be interpreted as solving a one-parameter family of Schr\"odinger EVPs (just as in eq.~(\ref{eq:standard_scattering_EVP}) above) defined by the scattering potentials \begin{align*}
    V(x;t) := \gamma u(x,t),
    \quad \text{with} \quad
    \gamma := -\frac{\alpha}{6\beta},
\end{align*} where $u = u(x,t)$ denotes the solution to the nonlinear PDE and $L(u) = -\pt^2_{\!x} \, + \, V(x;t)$ is now the Lax operator corresponding to the KdV equation. Here, the time derivative of the Lax operator is $ L_t = \gamma u_t$.

\newpage

For a fixed time $t \geq 0$, the \textit{direct scattering transform} $\mathcal{S}_t:u \mapsto \sigma(L(u)|_t)$ is defined as the nonlinear map which uniquely associates a field $u$ to its scattering data, \begin{align}\label{eq:DST}
    \mathcal{S}_t[u]
    := \sigma\big(L(u)|_t\big)
    = \big\{\big(\kappa_i(t), c_i(t)\big) \, : \, i=1, \dots,n\big\} \, \cup \, \big\{ \big(k, R(k,t)\big) \, : \, k \in \mathbb{R} \big\}.
\end{align} Conversely, the \textit{inverse scattering transform} $\mathcal{S}^{-1}_t:\sigma(L(u)) \mapsto u(\cdot,t)$ uses scattering data to reconstruct the field. In the context of the KdV-type models, the inverse transform $\mathcal{S}^{-1}_t$ can be implicitly defined for scattering data $\sigma$ of the form of eq.~(\ref{eq:scattering_data_def}) via \begin{align}\label{eq:IST}
    \vspace{-1mm}
    \mathcal{S}^{-1}_t[ \sigma ](x)
    := -\frac{2}{\gamma} \pt_{\!x} K_{\sigma}(x,x,t),
    \quad \text{with} \quad
    \mathcal{S}^{-1}_t\big[ \sigma\big(L(u)\big) \big](x) = u(x,t),
\end{align} where $K_{\sigma}$ is the unique interaction kernel satisfying the \textit{Gelfand-Levitan-Marchenko} (GLM) \textit{equation}, \begin{align}\label{eq:GLM_eqn}
    K_{\sigma}(x,y,t) \, + \, G_{\!\sigma}(x+y,t) \, + \int_x^{\infty} K_{\sigma}(x,x',t) \, G_{\!\sigma}(x' + y, t) \, dx' = 0,
    \quad \text{for} \quad
    y \geq x.
\end{align} Here, the function $G_{\!\sigma}$ denotes the observed reflection response and is defined by \begin{align*}
    G_{\!\sigma}(x;t) := \sum_{i=1}^{n} c_i(t) e^{ik_ix} + \frac{1}{2\pi} \int_{-\infty}^{\infty} R(k,t) e^{ikx} dk.
\end{align*} The GLM equation reconstructs a hidden 1D medium from its reflected waves, and thus can be thought of as a kind of deconvolution equation.

\smallskip

An important consequence of eq.~(\ref{eq:lax_evp}) is that the spectrum of $L$ is invariant in time; for any eigenpair $(\lambda, \phi)$ initially satisfying $L\phi = \lambda\phi$ at $t=0$, one has $\phi_t = M\phi$ and $\dot{\lambda} = 0$ for all $t \geq 0$ (see Appendix~\ref{appendixB:lax_pairs}).\\ For unperturbed systems (i.e., for $\epsilon = 0$), the scattering data also evolve linearly according to \begin{align}\label{eq:scattering_ODEs}
    \dot{R} = \boldsymbol{i}\omega(k)R,
    \quad \ \text{with} \ \quad
    \begin{cases}
        \dot{\kappa}_i = 0,
        \\
        \dot{c}_i = \omega(\kappa_i) c_i,
    \end{cases}
    \text{for each} \ \ i=1,\dots,n,
\end{align} where $\omega(k) := 8\beta k^3$ is the dispersion relation of the KdV equation~(\ref{eq:kdv_eqn}). The upshot is that the original nonlinear PDE can then be reduced to the system of linear ordinary differential equations (ODEs) in eq.~(\ref{eq:scattering_ODEs}) and solved via the following recipe: \begin{align*}
    u_0
    \, \xrightarrow{\text{DST}} \, \mathcal S_0[u]
    \, \xrightarrow{\text{integrate scattering ODEs}} \, \mathcal{S}_t[u]
    \, \xrightarrow{\text{IST}} u.
\end{align*} In the case that $0 < \epsilon \ll 1$, the adiabatic perturbation theory of Karpman and Solov’ev \cite{Karpman.Solovev1981PhysicaDNonlinearPhenomena} confirms that the equations of motion continuously deform to the tune of \begin{align}\label{eq:scattering_ODEs_perturbed}
    \dot{R} = \boldsymbol{i}\omega(k)R + \epsilon f_R(k) + \mathcal{O}(\epsilon^2),
    \quad \ \text{with} \ \quad
    \begin{cases}
        \dot{\kappa}_i = 0 + \epsilon f_{\kappa_i}(\boldsymbol{\kappa}, \mathbf{c}, R) + \mathcal{O}(\epsilon^2),
        \\
        \dot{c}_i = \omega(\kappa_i) c_i + \epsilon f_{c_i}(\boldsymbol{\kappa}, \mathbf{c}, R) + \mathcal{O}(\epsilon^2).
    \end{cases}
\end{align} A natural connection between the scattering ODEs given above and the governing PDE is obtained by recognizing that the field $u$ can be decomposed into $n$ localized waves parameterized in the fashion \begin{align}\label{eq:soliton_ansatz}
    u(x,t) = u_{\text{sol}}\big(x; \mathbf{x}_1(t), \dots, \mathbf{x}_n(t)\big) \, + \, u_{\text{rad}}(x,t),
\end{align} where $\mathbf{x}_i := (\kappa_i, c_i)$ represents the scattering data associated with the $i^{\rm{th}}$ soliton and $u_{\text{rad}}$ represents any excess \textit{radiation}. Importantly, for a pure-soliton solution ($u_{\text{rad}} = 0$), the system is \textit{reflectionless} ($R = 0$).

\newpage

\subsection{Sparse Regression with WSINDy}\label{subsec:wsindy}
Consider a state vector $\boldsymbol{x}(t) = (x_1, \dots, x_d)(t) \in \mathbb{R}^d$ governed by a system of ODEs\footnote{In this paper, we will be concerned with ordinary differential equations. However, the SINDy \cite{Brunton.etal2016ProcNatlAcadSciUSA} and WSINDy \cite{Messenger.Bortz2021MultiscaleModelSimul} paradigms can also be extended to work with partial differential equations; see \cite{Messenger.Bortz2021JournalofComputationalPhysics, Rudy.etal2017SciAdv, schaefferLearningPartialDifferential2017}.} of the form \begin{align}\label{eq:sindy_ODE}
    \dot{\boldsymbol{x}} = \mathbf{\Theta}(\boldsymbol{x})\mathbf{w},
    \quad \text{with} \quad
    \begin{cases}
        \mathbf{\Theta}(\boldsymbol{x}) := [f_1(\boldsymbol{x}), \, \dots \, , \, f_J(\boldsymbol{x})] \in \mathbb{R}^{J},
        \\
        \mathbf{w} = [\mathbf{w}_1, \, \dots \, , \, \mathbf{w}_d] \in \mathbb{R}^{J \times d},
    \end{cases}
\end{align} where each $f_j:\mathbb{R}^d \rightarrow \mathbb{R}$ represents a scalar-valued function of the state, given component-wise by \begin{align*}
    \dot{x}_i = \mathbf{\Theta}(\boldsymbol{x})\mathbf{w}_i,
    \quad \text{for each} \quad
    i = 1, \dots, d.
\end{align*} The \textit{weak-form sparse identification of nonlinear dynamics} (WSINDy) algorithm \cite{Messenger.Bortz2021JournalofComputationalPhysics} is a data-driven technique which attempts to infer dynamics of the form of eq.~(\ref{eq:sindy_ODE}) above from noisy training data \begin{align*}
    \mathcal{X} = \big\{\boldsymbol{x}(t_m) + \boldsymbol{\varepsilon}(t_m) \, : \, m = 1, \dots, N_t\big\},
    \quad \text{where, e.g.,} \quad
    \boldsymbol{\varepsilon}(t) \sim \mathcal{N}\big(0, \sigma^2_{\!\varepsilon}\mathbf{I}\big).
\end{align*} This is accomplished by integrating the governing ODE against a collection of smooth and compactly supported test functions, \begin{align*}
    \varphi_k(t) := \varphi(t_{m_k} - t),
    \quad \text{for} \ \ \ k = 1, \dots, K,
\end{align*} where $\varphi \in C^{\infty}_c$ is symmetric about $t=0$ and the \textit{query points}, $t_{m_k} \in [0,T]$, are uniformly placed within the domain. Using integration by parts, models of the form of eq.~(\ref{eq:sindy_ODE}) are then transformed into their \textit{convolutional weak formulations},\footnote{Note that the factor of `$-1$' resulting from integration by parts is eliminated by adopting the sign convention $(\varphi * f)(t_k) = \langle \varphi(t_{m_k} - \cdot), f(\cdot)\rangle$.} \begin{align}\label{eq:weak_convolutional_form}
    \big(\dot{\varphi} * \boldsymbol{x}\big)(t_{m_k}) = \big(\varphi * \mathbf{\Theta}(\boldsymbol{x}) \mathbf{w}\big)(t_{m_k}).
\end{align} Accordingly, we investigate the linear system given by \begin{align}\label{eq:weak_linear_system}
    \mathbf{b}(\boldsymbol{x}) = \mathbf{G}(\boldsymbol{x}) \mathbf{w},
    \quad \text{where} \quad
    \begin{cases}
        \mathbf{b}(\boldsymbol{x})_{k,i} := \big(\dot{\varphi} * x_i\big)(t_{m_k}) = \langle \dot{\varphi}_k, x_i \rangle,
        \\
        \mathbf{G}(\boldsymbol{x})_{k,j} := \big(\varphi * f_j(\boldsymbol{x})\big)(t_{m_k}) = \langle \varphi_k, f_j(\boldsymbol{x}) \rangle,
    \end{cases}
\end{align} where $\langle\cdot\,,\cdot\rangle$ denotes the $L^2$ inner-product. Observe that in eqs.~(\ref{eq:weak_convolutional_form}) and (\ref{eq:weak_linear_system}), no differential operators are applied directly to the state variable $\boldsymbol{x} = \boldsymbol{x}(t)$ -- a property which increases the fidelity of the results in noisy regimes (that is, compared to the strong-form of SINDy \cite{Brunton.etal2016ProcNatlAcadSciUSA}; see, e.g., Table 6 in \cite{Messenger.Bortz2021JournalofComputationalPhysics}).

\smallskip

In contrast to standard parameter estimation problems, here the form of the governing equation is \textit{not} assumed to be known prima-facie. Instead, one typically assumes in the context of sparse regression that the \textit{library} of candidate terms $\mathbf{\Theta}(\boldsymbol{x})$ is substantially over-specified, and that only a small subset of its parameters are nonzero (i.e., the parameter matrix $\mathbf{w}$ is sparse). In turn, WSINDy targets a regularized least-squares problem of the form \begin{align}\label{eq:sparse_regression_problem}
    \hat{\mathbf{w}} := \argmin_{\mathbf{w} \in \mathbb{R}^{J \times d}} \, \mathcal{L}_{\mu}(\mathbf{w}),
    \quad \text{where} \quad
    \mathcal{L}_{\mu}(\mathbf{w}) := \left\|\mathbf{b}(\boldsymbol{x}) - \mathbf{G}(\boldsymbol{x})\mathbf{w}\right\|^2_F + \mu\|\mathbf{w}\|_0.
\end{align} The regularization term $\mu \|\mathbf{w}\|_0$ in the loss function promotes the selection of a parsimonious model by penalizing the number of nonzero coefficients, where $\|\cdot\|_0$ denotes the $\ell_0$ pseudo-norm. Since the $\ell_0$ penalty is non-differentiable, in practice the loss function in eq.~(\ref{eq:sparse_regression_problem}) is approximately minimized via iterative thresholding schemes, e.g., \textit{modified sequential thresholding least squares} (MSTLS) \cite{Brunton.etal2016ProcNatlAcadSciUSA, Messenger.Bortz2021JournalofComputationalPhysics}, which progressively restricts the columns of $\boldsymbol{\Theta}(\boldsymbol{x})$ available to the model (see \S\ref{appendixA:MSTLS} of the Appendix).

\newpage

\section{Methods}\label{sec:methods}
\subsection{High-Level Overview}\label{subsec:outline}
As briefly mentioned in \S\ref{subsec:previous_work}, Yang et al. have in recent work \cite{Yang.etal2025} explored data-driven soliton dynamics (specifically, in the context of the nonlinear Schr\"odinger equation) by explicitly fitting parameterized waveforms to data. In particular, these authors developed a pipeline of the form
\begin{align*}
    \underbrace{\big\{ u(x,t) \big\}}_{\text{measure data}}
    \quad \rightarrow \quad
    \underbrace{\big\{ \boldsymbol{x}_i = (\xi_i, a_i, v_i, \theta_i) \big\}_{i=1}^{n}}_{\text{extract fit parameters}}
    \quad \rightarrow \quad
    \underbrace{\dot{\boldsymbol{x}}_i \approx \mathbf{\Theta}(\boldsymbol{x}) \hat{\mathbf{w}}_i, \ \ \text{for} \ \ i = 1, \dots, n}_{\text{infer effective ODEs with SINDy}},
\end{align*} where $\xi_i$, $a_i$, $v_i$, and $\theta_i$ respectively denote the centers, amplitudes, velocities, and phases of modulated $\text{sech}$-like profiles fit to the $i^{\rm{th}}$ soliton. For the reasons detailed in \S\ref{subsec:previous_work}, we propose an alternative pipeline: \begin{align*}
    \underbrace{\big\{ u(x,t) \big\}}_{\text{measure data}}
    \quad \rightarrow \quad
    \underbrace{\big\{ \boldsymbol{x}_i = (\kappa_i, c_i) \big\}_{i=1}^{n}}_{\text{compute scattering data}}
    \quad \rightarrow \quad
    \underbrace{\mathbf{b}(\boldsymbol{x}_i) \approx \mathbf{G}(\boldsymbol{x}) \hat{\mathbf{w}}_i, \ \ \text{for} \ \ i = 1, \dots, n}_{\text{infer effective ODEs with WSINDy}}.
\end{align*} Given our focus on effective models for solitons (and to simplify the core ideas of the paper), in this work we specifically address reflectionless settings with $R = 0$; however, in the regime of nonzero radiation, note that one can simply append additional components of the form $R(\mathbf{k},t) \in \mathbb{R}^{N_{\mathbf{k}}}$ onto the state vector.

\subsection{Numerical Implementation}\label{subsec:numerical_implementation}
\begin{figure}[htb]
    \centering
    \!\!\!\!\includegraphics[width=1.02\linewidth]{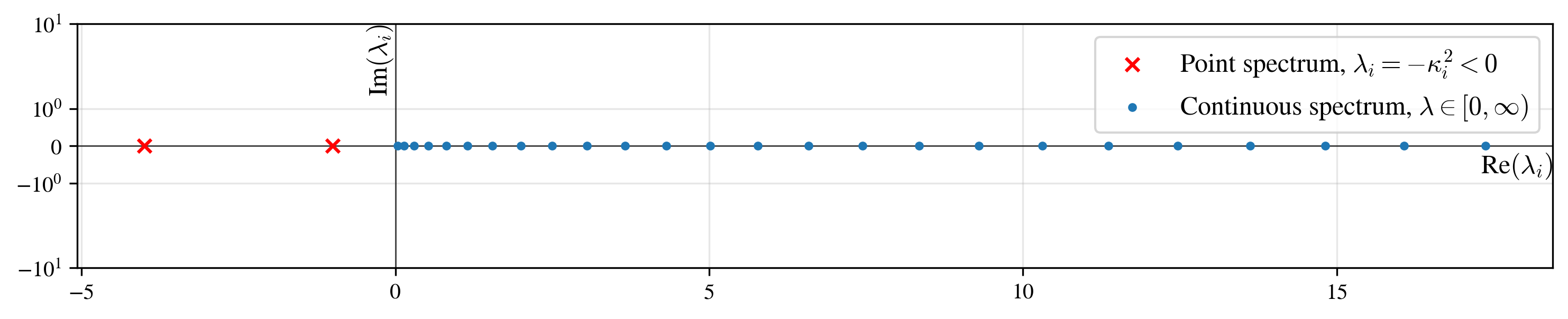}
    \\
    \includegraphics[width=\linewidth]{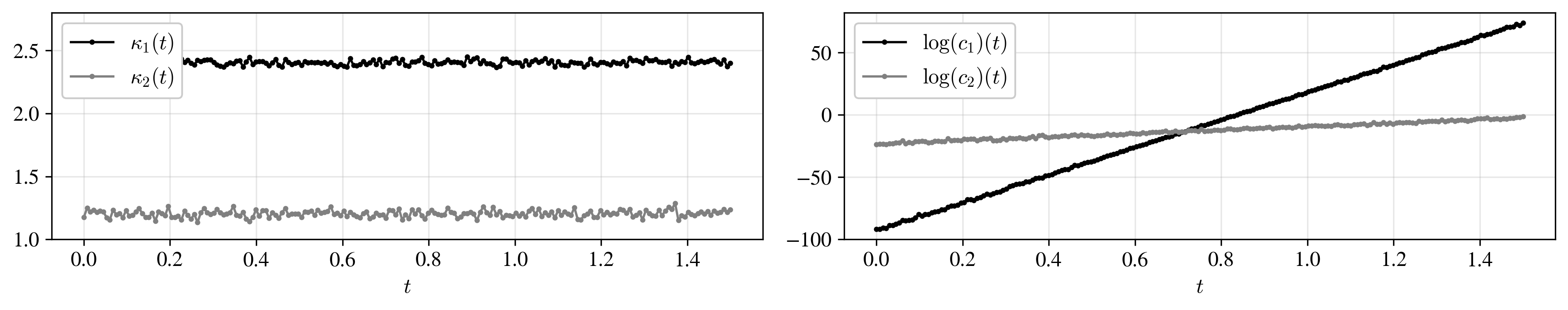}
    \vspace{-4.9mm}
    \\
    \includegraphics[width=0.96\linewidth]{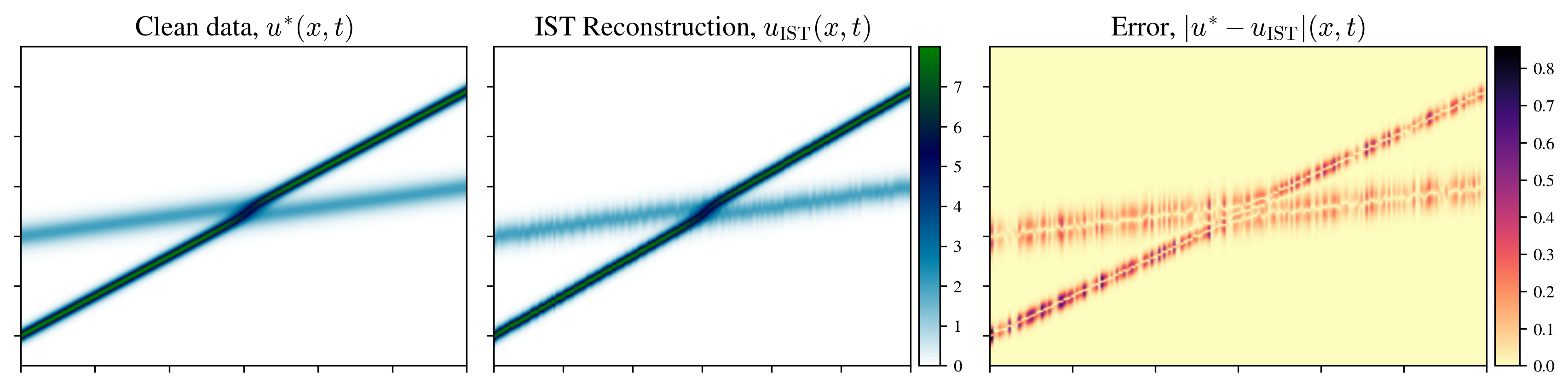}
    \caption{\normalsize Illustrating the numerical computation of the scattering data, in this case using the noisy measurements from the two-soliton collision example of \S\ref{subsec:synthetic_data} (see Figure~\ref{fig:kdv_bright_soliton_snapshots}). [Top panel] Plotting the first few eigenvalues in the time-invariant spectrum of the discretized Lax operator $\mathbf{L}(\mathbf{u}^*)$, for reference; the `bound state' eigenvalues $\lambda_{i} = -\kappa^2_i < 0$ in the point spectrum parameterize the soliton dynamics. [Middle panel] Plotting the numerically computed $\kappa_i(t)$ and $\log(c_i)(t)$ time-series; note that these quantities correlate with the soliton widths and locations, respectively. [Bottom panel] Comparing the true state $\mathbf{u}^*(t)$ to the reconstructed state $\mathbf{u}_{\text{IST}}(t)$ obtained by applying the IST to these scattering data}
    \label{fig:scattering_data}
\end{figure}

In our numerical implementation, we discretize the state vector $\boldsymbol{x}(t) \in \mathbb{R}^d$ over a uniform temporal grid $\mathbf{t} = [t_1, \dots, t_{N_t}]^T \subset [0,T]$ with spacing $\Delta{t}$, producing a data matrix $\mathbf{X} := \mathbf{X}^* + \boldsymbol{\epsilon}(\mathbf{t}) \in \mathbb{R}^{N_t \times \, d}$ given by \begin{align*}
    \mathbf{X}
    = \begin{bmatrix}
        x_1(t_1) & \cdots & x_d(t_1) \\
        \vdots & & \vdots \\
        x_1(t_{N_t}) & \cdots & x_d(t_{N_t})
    \end{bmatrix}
    = \begin{bmatrix}
        x^*_1(t_1) & \cdots & x^*_d(t_1) \\
        \vdots & & \vdots \\
        x^*_1(t_{N_t}) & \cdots & x^*_d(t_{N_t})
    \end{bmatrix}
    +
    \begin{bmatrix}
        \epsilon_1(t_1) & \cdots & \epsilon_d(t_1) \\
        \vdots & & \vdots \\
        \epsilon_1(t_{N_t}) & \cdots & \epsilon_d(t_{N_t})
    \end{bmatrix},
\end{align*} where $\mathbf{X}^* := \boldsymbol{x}^*(\mathbf{t})$ denotes the matrix of uncorrupted (i.e., non-noisy) data. The data matrix $\mathbf{X}$ represents the scattering data computed from observations of a noisy scalar field $\mathbf{u}(t) \in \mathbb{R}^{N_x}$ of the form \begin{align*}
    \mathbf{u}(t) := \mathbf{u}^*(t) + \boldsymbol{\varepsilon}(t),
    \quad \text{where} \quad
    \mathbf{u}(t) := \texttt{vec}\big\{u(n\Delta{x},t) \, : \, n\Delta{x} \in \Omega\big\},
\end{align*} and a discretized Lax operator $\mathbf{L}(\mathbf{u}) \in \mathbb{R}^{N_x \times N_x}$ defined by \begin{align*}
    \mathbf{L}(\mathbf{u}) := -\mathbf{D}_{xx} + \gamma\mathbf{u}' = \mathbf{Q} \mathbf{\Lambda} \mathbf{Q}^T\!,
    \quad \text{with} \quad
    (\,\cdot\,)' := \texttt{diag}(\,\cdot\,),
\end{align*} where $\mathbf{D}_{xx}$ is a second-order centered finite difference operator and $\mathbf{Q}, \mathbf{\Lambda}$ give a unitary diagonalization of $\mathbf{L}(\mathbf{u})$. After ordering the entries of $\mathbf{\Lambda}$ in ascending order, the first $n$ negative eigenvalues approximate the point spectrum of $L(u)$, and we thus define $\lambda_i = -\kappa^2_i := \Lambda_{ii} < 0$ and $\phi_i(\mathbf{x},t) := \mathbf{q}_i(t)$ for $i = 1, \dots, n$, where $\mathbf{Q} = [\mathbf{q}_1, \dots, \mathbf{q}_{N_x}]$; see Figure~\ref{fig:scattering_data}. The discrete scattering data then take the form \begin{align*}
    \boldsymbol{x}(t) := \mathcal{S}_t[\mathbf{u}] = [\kappa_1(t), \dots, \kappa_n(t), c_1(t), \dots, c_n(t)] \in \mathbb{R}^d,
    \quad \text{where} \quad
    d := 2n,
\end{align*} and where the norming constants $c_i$ are computed as per eq.~(\ref{eq:norming_constant}) after numerically solving the boundary value problem in eq.~(\ref{eq:jost_soln}) for the Jost solution.


\newpage

It is important to recognize that the DST map $u \mapsto \mathcal{S}_t[u]$ is highly nonlinear and, as such, can be sensitive to noisy perturbations of the form $u = u^* + \varepsilon$. Although this presents limitations when working with experimental data, this sensitivity is partially alleviated by working within the noise-robust weak formulation described in \S\ref{subsec:wsindy}. Letting $\mathbf{Q}^*, \mathbf{\Lambda}^*$ provide a unitary diagonalization of $\mathbf{L}(\mathbf{u}^*)$, one finds that \begin{align*}
    \mathbf{L}(\mathbf{u}) = \mathbf{L}(\mathbf{u}^*) + \gamma\boldsymbol{\varepsilon}',
    \quad \text{or equivalently,} \quad
    \mathbf{Q \Lambda Q}^T = \mathbf{Q}^*\mathbf{\Lambda}^*{\mathbf{Q}^*}^T + \gamma\boldsymbol{\varepsilon}'.
\end{align*} Applying first-order eigenvalue perturbation theory, one in turn finds that the eigenvalues are perturbed to the tune of \begin{align*}
    \lambda_i = \lambda^*_{i} \, + \, \gamma\|\mathbf{q}_i\|^2_{\boldsymbol{\varepsilon}'} + \, \mathcal{O}\big(\|\mathbf{q}_i\|^4_{\boldsymbol{\varepsilon}'}\big),
\end{align*} meaning that, to first-order, $\lambda_i = -\kappa^2_i$ is approximately normally-distributed about $\lambda_i = -(\kappa^*_i)^2$ with a standard deviation that is proportional to the weighted $L^2$ norm $\|\mathbf{q}_i\|^2_{\boldsymbol{\varepsilon}'}$. Similar results also hold for the norming constants $c_i$, which implicitly depend on the sensitivity of the Jost solution $\phi_{+}$ to measurement error. Adding noise can also result in the generation of spurious small negative eigenvalues, which in practice we find can be avoided by numerically enforcing a lower-bound of the form $\kappa_{\min} \in [0.2, 0.6]$.

\clearpage

We discretize the variational WSINDy problem posed in eq.~(\ref{eq:weak_linear_system}) by reshaping the candidate library matrix to the tune of $\mathbf{\Theta}(\boldsymbol{x}) \in \mathbb{R}^J \mapsto \mathbf{\Theta}(\mathbf{X}) \in \mathbb{R}^{M \times J}$, where \begin{align*}
    \mathbf{\Theta}(\mathbf{X})
    := \begin{bmatrix}
        f_1(\boldsymbol{x})(t_1) & \cdots & f_J(\boldsymbol{x})(t_1) \\
        \vdots & & \vdots \\
        f_1(\boldsymbol{x})(t_{N_t}) & \cdots & f_J(\boldsymbol{x})(t_{N_t})
    \end{bmatrix}.
\end{align*} In turn, banded convolution matrices $\mathbf{\Phi}, \dot{\mathbf{\Phi}} \in \mathbb{R}^{K \times {N_t}}$ can be defined via \begin{align*}
    \mathbf{\mathbf{\Phi}} :=
    \begin{bmatrix}
        \varphi_1(t_1) & \cdots & \varphi_1(t_{N_t}) \\
        & \ddots & \\
        \varphi_K(t_1) & \cdots & \varphi_K(t_{N_t})
    \end{bmatrix}
    \quad \text{and} \quad
    \dot{\mathbf{\mathbf{\Phi}}} :=
    \begin{bmatrix}
        \dot{\varphi}_1(t_1) & \cdots & \dot{\varphi}_1(t_{N_t}) \\
        & \ddots & \\
        \dot{\varphi}_K(t_1) & \cdots & \dot{\varphi}_K(t_{N_t})
    \end{bmatrix},
    \quad \text{with} \ \ \,
    \begin{cases}
        \mathbf{b}(\mathbf{X}) := \dot{\mathbf{\mathbf{\Phi}}} \mathbf{X},
        \\
        \mathbf{G}(\mathbf{X}) := \mathbf{\mathbf{\Phi}} \mathbf{\Theta}(\mathbf{X}).
    \end{cases}
\end{align*} The discretized analogue of the weak convolutional formulation in eq.~(\ref{eq:weak_convolutional_form}) is then given, up to a quadrature error $\|\mathbf{e}_{\text{int}}\| = \mathcal{O}(\Delta{t}^{p+1})$ with $2p$ being the polynomial order of $\varphi$, by $\mathbf{b}(\mathbf{X}^*) \approx \mathbf{G}(\mathbf{X}^*)\mathbf{w}^*$. Similarly, the discrete analogue of the sparse regression problem in eq.~(\ref{eq:sparse_regression_problem}) takes the form \begin{align}\label{eq:discrete_weak_form_loss}
    \hat{\mathbf{w}} := \argmin_{\mathbf{w} \in \mathbb{R}^{J \times d}} \, \mathcal{L}_{\mu}(\mathbf{w}),
    \quad \text{where} \quad
    \mathcal{L}_{\mu}(\mathbf{w}) := \left\|\mathbf{b}(\mathbf{X}) - \mathbf{G}(\mathbf{X})\mathbf{w}\right\|^2_F + \mu\|\mathbf{w}\|_0,
\end{align} the solution of which we approximate via the MSTLS algorithm described in Appendix~\ref{appendixA:MSTLS}.


\subsection{Test Functions and Candidate Library}\label{subsec:candidate_library}
Following \cite{Messenger.Bortz2021JournalofComputationalPhysics}, we use a rescaled Bernstein polynomial for the generating test function $\varphi$, defined in a piecewise fashion by \begin{align*}
    \varphi(t;m,p) := \left(1 - \frac{t^2}{m^2\Delta{t}^2} \right)^{p}
    \quad \text{within} \quad
    t \in \text{supp}(\varphi) := \, [-m\Delta{t}, \, m\Delta{t}],
\end{align*} where the degree $p$ is defined for a highest derivative order $\bar{\alpha} := 1$ and support tolerance $\tau_0 := 1\texttt{e}-10$ via \begin{align}\label{eq:test_fcn_degree}
    p = \max\left\{ \left\lceil
    \frac{\ln(\tau_0)}{\ln((2m-1)/m^2)}
    \right\rceil\!, \ \bar{\alpha} + 1 \right\} \geq 2.
\end{align} Convolving the scattering data against localized test function kernels $\varphi_k(\cdot;m,p)$ approximately projects these data onto the temporal scales $t \gtrapprox m\Delta{t}$; in particular, as the support radius $m\Delta{t} \rightarrow 0$, the weak (WSINDy) formulation in eq.~(\ref{eq:weak_convolutional_form}) collapses to the strong (SINDy) formulation in eq.~(\ref{eq:sindy_ODE}). From a distributional perspective, the test functions $\varphi_k$ converge to Dirac delta distributions $\delta(t_k)$ while the derivatives $\dot{\varphi}_k$ converge to centered finite difference kernels. We provisionally set $m := 20$ throughout. For additional information about our numerical implementation, we refer the reader to Appendix~\S\ref{appendixA:MSTLS}.

\smallskip

Our choice of library $\mathbf{\Theta}(\boldsymbol{x})$ is motivated by the analytic structure of the scattering ODEs in eq.~(\ref{eq:scattering_ODEs}). It is convenient to work in terms of $(\kappa_i,\log(c_i))$ rather than, e.g., $(\kappa_i,c_i)$ or $(\lambda_i, c_i)$, as the scattering dynamics of the KdV equation become polynomial in these coordinates: $\dot\kappa_i=0$ and $\tfrac{d}{dt}\log(c_i) = 8\beta\kappa_i^3$. Accordingly, we use a library of low-order (i.e., cubic) monomials in $\kappa_i$ and in
$\log(c_i)$,
\begin{equation}\label{eq:candidate_library}
    \mathbf{\Theta}(\boldsymbol{x}_i) = \big\{\kappa_i,\, \kappa^2_i,\, \kappa^3_i, \, \log(c_i), \, \log(c_i)^2, \, \log(c_i)^3\big\}.
\end{equation} In order to respect the index-invariant physical symmetry of the system -- i.e., that each soliton follows the same physics -- we enforce index-invariant ODEs of the form $\dot{\kappa}_i = f_{\boldsymbol{\kappa}}(\kappa_i, c_i)$ and $\dot{c}_i = f_{\mathbf{c}}(\kappa_i, c_i)$.


\newpage

\section{Results}\label{sec:results}

\subsection{Validation}\label{subsec:validation}
To illustrate how the performance of our scheme scales with increasingly noisy data $\mathbf{u} = \mathbf{u}^* + \boldsymbol{\varepsilon}$, we run repeated trials on corrupted versions of the datasets and report the resulting metrics (see Figure~\ref{fig:validation_plots}). Following \cite{Messenger.Bortz2021JournalofComputationalPhysics}, we add distinct realizations of artificial i.i.d. noise $\varepsilon \in \mathcal{N}(0,\sigma^2_{\!\varepsilon})$ in a pointwise fashion to each element of $\mathbf{U} := \mathbf{u}(\mathbf{t})$, which are computed by enforcing a standard deviation $\sigma_{\!\varepsilon} = \sigma_{\textsc{nr}} \| \mathbf{U} \|_F$ so that $\sigma_{\textsc{nr}} = \| \boldsymbol{\varepsilon}(\mathbf{t}) \|_F / \| \mathbf{U} \|_F$. We explore noise ratios in the range $0 \leq \sigma_{\textsc{nr}} \leq 0.5$, i.e., up to $50\%$ of the magnitude of the data. To give a sense of the symbolic form of the discovered equations, in Table~\ref{table:kdv-synthetic-results} we explicitly compare the ground-truth and identified models at zero noise for the synthetically generated KdV data of \S\ref{subsec:synthetic_data}. Metrics given below then track how the identified models deform as $\sigma_{\textsc{nr}}$ increases.

\smallskip

To help gauge the quality of the weak-form regression, we report the coefficient of determination corresponding to each identified WSINDy model, which is defined by
\footnote{The $R^2$ metric is unstable for models with small or vanishing coeffs; therefore, we do not report an $R^2$ value when $\hat{\mathbf{w}}=0$ (see Figure~\ref{fig:validation_plots}).}
\begin{align*}
    R^2 := 1 - \frac{\| \mathbf{b} - \mathbf{G}\hat{\mathbf{w}} \|_2^2}{\| \mathbf{b} - \overline{\mathbf{b}} \, \|_2^2},
    \quad \text{where} \quad
    \overline{\mathbf{b}} := \left[\frac{1}{Kd} \sum_{k,i} \mathbf{b}_{k,i}\right] \mathbf{1}.
\end{align*} This metric, which equals the proportion of the variance of $\mathbf{b}$ that is explained by the identified model $\mathbf{G}\hat{\mathbf{w}}$, satisfies $R^2 \leq 1$, with values near one indicating an accurate model. We additionally report the pointwise RMSE between the true state $\mathbf{U}^*$ and the reconstructed field $\mathbf{U}_{\text{IST}}$ obtained by applying the IST to a forward simulation of the discovered scattering dynamics (see the bottom panel of Figure~\ref{fig:scattering_data}), \begin{align}\label{eq:IST_reconstruction_RMSE}
    \text{RMSE}(\mathbf{U}_{\text{IST}})
    := \frac{1}{\sqrt{N_{\!x}N_t}} \big\| \mathbf{U}^* - \mathbf{U}_{\text{IST}} \big\|_F,
\end{align} which quantifies how faithfully the recovered dynamics reproduce the original field. For reflectionless data, the IST admits a convenient closed-form "$\tau$-determinant" representation \cite{Hirota1971PhysRevLett} given by \begin{align*}
    u(x,t) = -\frac{2}{\gamma} \frac{d^2}{dx^2} \log|\mathbf{A}(x; \boldsymbol{\kappa}(t), \mathbf{c}(t))|,
    \quad \text{where} \quad
    \mathbf{A}_{ij}(x; \boldsymbol{\kappa}, \mathbf{c}) := \delta_{ij} + \frac{\sqrt{c_ic_{\!j}}\, e^{-(\kappa_i + \kappa_j)x}}{\kappa_i + \kappa_j},
\end{align*} which we use to recover $\mathbf{U}_{\text{IST}}$ from the scattering data $(\hat{\kappa}_i, \hat{c}_i)(t)$ forecast with the identified models.\footnote{The $\tau$-determinant representation is more numerically convenient than the GLM equation of eq.~(\ref{eq:GLM_eqn}), but only applies to reflectionless cases. When forecasting a WSINDy model forward in time, we use an adaptive RK-45 scheme instantiated with the exact initial condition; see Figure~\ref{fig:heinrich_et_al_data}.}

\smallskip

The metrics above are model-agnostic and can be computed in every case, including the experimental setting of \S\ref{subsec:empirical_data}. In cases where underlying the model is available, i.e., the synthetic experiments of \S\ref{subsec:synthetic_data}, we also compare the identified coefficients against a reference; for unperturbed synthetic data ($\epsilon = 0$) the reference is the exact scattering ODEs of eq.~(\ref{eq:scattering_ODEs}), while for perturbed synthetic data ($\epsilon > 0$) it is the perturbed scattering ODEs of eq.~(\ref{eq:scattering_ODEs_perturbed}). In these cases, we follow \cite{Messenger.Bortz2021JournalofComputationalPhysics} in reporting the normalized $\ell^{\infty}$ coefficient error and the true positive ratio (TPR),\footnote{Here, $\text{TP}$ denotes the number of terms that were correctly identified as nonzero, $\text{FP}$ denotes the number of terms that were falsely identified as nonzero, and $\text{FN}$ denotes the number of terms that were falsely identified as zero.} respectively defined by \begin{align}\label{eq:Einfty_and_TPR}
    E_{\infty} := \max_{j=1,\dots,J} \frac{|w_j - w^*_j|}{|w^*_j|},
    \quad \text{and} \quad
    \text{TPR} := \frac{\text{TP}}{\text{TP} + \text{FP} + \text{FN}}.
\end{align} The $E_{\infty}$ coefficient error represents maximum element-wise relative error incurred by the discovered model, while the TPR assesses the extent to which the identified models recover the correct terms. Note that a TPR of 1 means that the true model has been discovered in its entirety while a TPR of 0 indicates that none of the correct terms were identified.

\clearpage
\begin{landscape}

\begin{table}[p]
\TBL{
\caption{Results obtained using synthetic, noise-free ($\sigma_{\textsc{nr}} = 0$) data sourced from numerical simulations of the perturbed KdV model in eq.~(\ref{eq:perturbed_kdv_eqn}) with the canonical parameters $(\alpha,\beta) = (6,1)$. For each configuration, ground-truth reference dynamics are compared against the model identified by WSINDy. The reference equations are given by the exact scattering dynamics of eq.~(\ref{eq:scattering_ODEs}) for the unperturbed runs ($\epsilon = 0$) and by the near-integrable scattering dynamics of eq.~(\ref{eq:scattering_ODEs_perturbed}) for the perturbed runs ($\epsilon = 0.2$). For $\epsilon = 0.2$, note that $\frac{1}{3}\epsilon \approx 0.0667$ and $\frac{2}{3}\epsilon \approx 0.1333$}
\label{table:kdv-synthetic-results}
}
{%
\small
\renewcommand{\arraystretch}{1.55}

\begin{tabular*}{\linewidth}{@{\extracolsep{\fill}}%
L{0.11\linewidth} L{0.11\linewidth} L{0.15\linewidth}
L{0.30\linewidth} L{0.30\linewidth}@{}}

\toprule
\TCH{\shortstack{Number of\\solitons $(n)$}} &
\TCH{\shortstack{Perturbation $(\epsilon)$}} &
\TCH{\shortstack{Forcing $(F)$}} &
\TCH{Ground truth model} &
\TCH{Identified model $(0\%$ noise)}
\\
\midrule

\resultrow
{$n=1$}
{$\epsilon=0$}
{N/A}
{\resultcell{\dot{\kappa} = 0, \\ \frac{d}{dt}\log(c) = 8\kappa^3}}
{\resultcell{\dot{\kappa} = {\color{purple}0}, \\ \frac{d}{dt}\log(c) = {\color{purple}7.9981}\kappa^3}}

\lightdashline

\resultrow
{$n=2$}
{$\epsilon=0$}
{N/A}
{\resultcell{\dot{\kappa}_i = 0, \\ \dfrac{d}{dt}\log(c_i) = 8\kappa_i^3}}
{\resultcell{\dot{\kappa}_i = {\color{purple}0}, \\ \frac{d}{dt}\log(c_i) = {\color{purple}7.9984}\kappa_i^3}}

\midrule

\resultrow
{$n=1$}
{$\epsilon=0.2$}
{$F[u]=(xu)_x$}
{\resultcell{\dot{\kappa} = \frac{1}{3}\epsilon\kappa + O(\epsilon^2), \\ \frac{d}{dt}\log(c) = 8\kappa^3 - \frac{2}{3}\epsilon\log(c) + O(\epsilon^2)}}
{\resultcell{\dot{\kappa} = {\color{purple}0.0666}\kappa, \\ \frac{d}{dt}\log(c) = {\color{purple}8.0072}\kappa^3 {\color{purple}- 0.1336}\log(c)}}

\lightdashline

\resultrow
{$n=2$}
{$\epsilon=0.2$}
{$F[u]=(xu)_x$}
{\resultcell{\dot{\kappa}_i = \frac{1}{3}\epsilon\kappa_i + O(\epsilon^2), \\ \frac{d}{dt}\log(c_i) = 8\kappa^3_i - \frac{2}{3}\epsilon\log(c_i) + O(\epsilon^2)}}
{\resultcell{\dot{\kappa}_i = {\color{purple}0.0660}\kappa_i, \\ \frac{d}{dt}\log(c_i) = {\color{purple}8.0192}\kappa^3_i {\color{purple}- 0.1364}\log(c_i)}}

\botrule
\end{tabular*}}
\end{table}
\end{landscape}
\clearpage

\begin{figure}[htb]
    \centering
    \includegraphics[width=\linewidth]{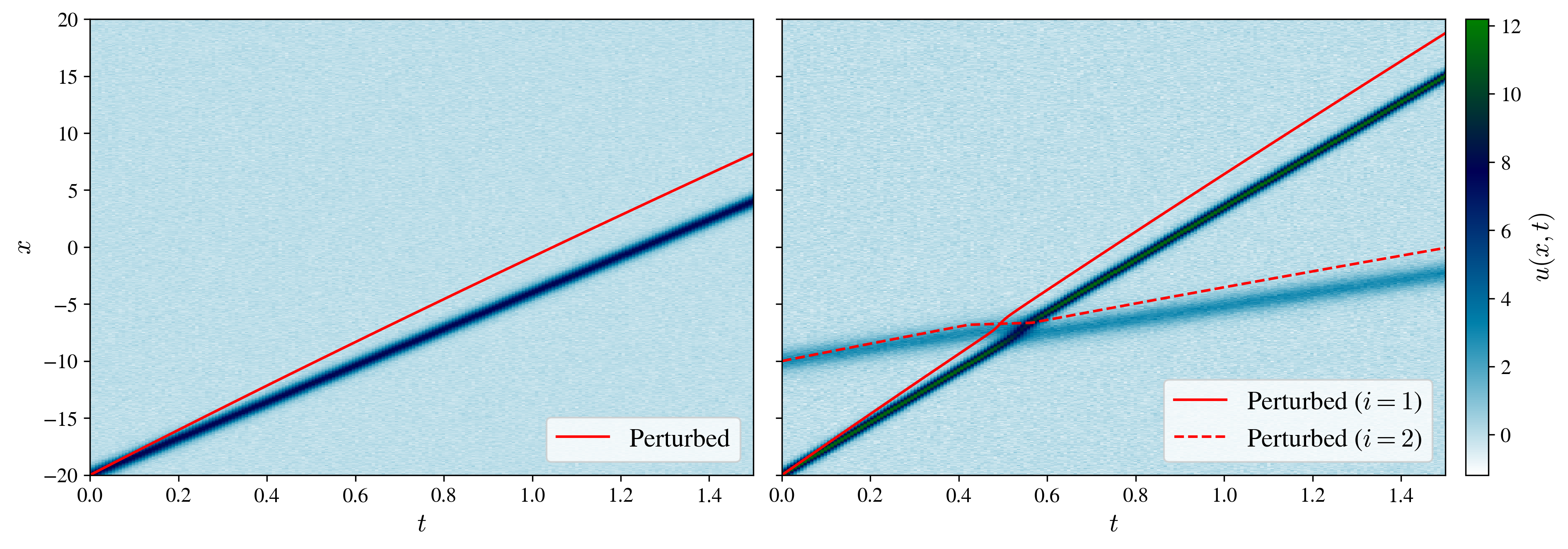}
    \\
    \includegraphics[width=0.995\linewidth]{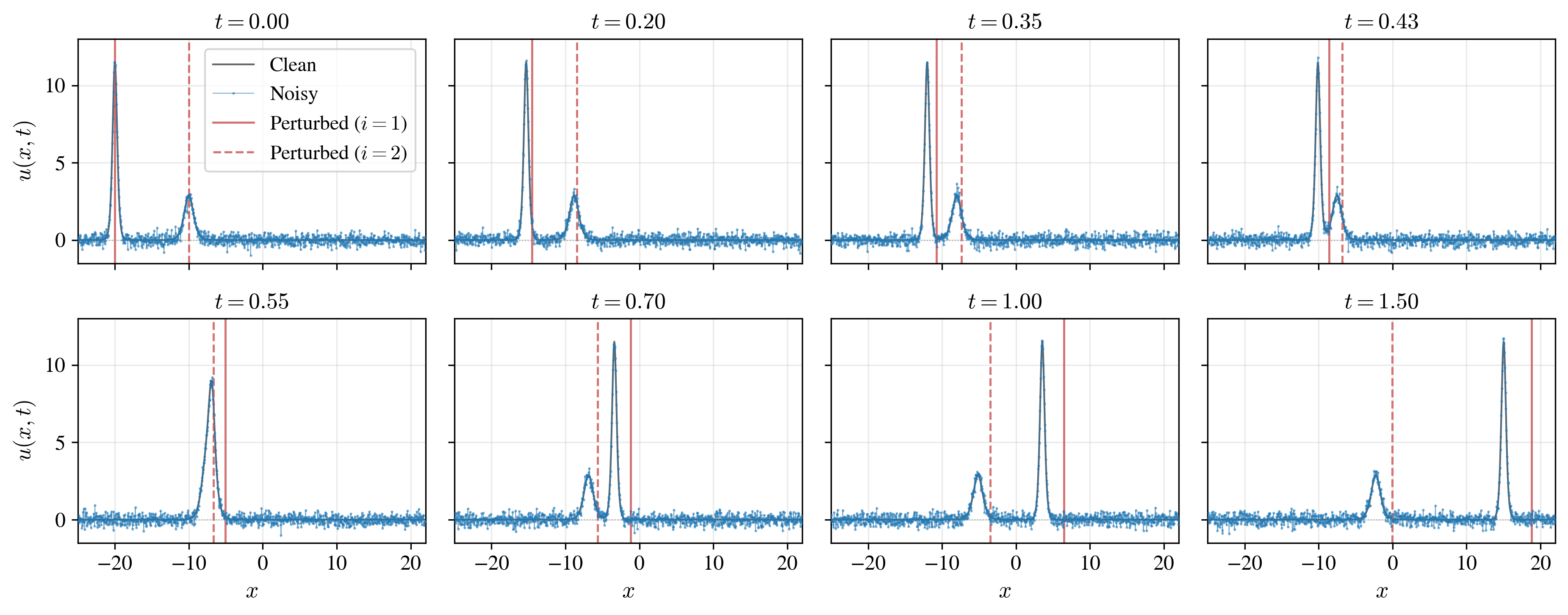}
    \caption{[Top panel] Visualizing one of three sets of numerical simulations used to generate synthetic data, which each follow the evolution of one or two solitons evolving under the perturbed KdV equation in eq.~(\ref{eq:perturbed_kdv_eqn}), pictured at $20\%$ noise with $\kappa=2$ and $(\kappa_1,\kappa_2)=(2.4, 1.2)$ in the left and right panels, respectively. [Bottom panel] Snapshots of the double-soliton simulations evolving in time}
    \label{fig:kdv_bright_soliton_snapshots}
\end{figure}

\vspace{-6mm}

\subsection{Synthetic Data}\label{subsec:synthetic_data}
We consider several examples using synthetic data sourced from numerical simulations of the perturbed KdV equation given in eq.~(\ref{eq:perturbed_kdv_eqn}). In particular, we explore configurations featuring either one or two solitons (see Figure~\ref{fig:kdv_bright_soliton_snapshots}) and investigate both the unperturbed ($\epsilon = 0$) and perturbed cases ($\epsilon = 0.2$) subject to a mass-conserving forcing function, $F[u] = (xu)_x$, for which the reference ODEs in eq.~(\ref{eq:scattering_ODEs_perturbed}) can be explicitly computed (see Table~\ref{table:kdv-synthetic-results}). In each instance, we use the canonical KdV parameters $\alpha = 6$ and $\beta = 1$. The numerical simulations are computed on the domain $x \in \Omega = [-25,25]$ using $N = 2^{10}$ uniformly-spaced nodes and adaptively integrated for $t \in [0,1.5]$ using a pseudo-spectral solver from the open-source $\texttt{sangkuriang-ideal}$ \cite{Irawan.etal2026} Python package with $M=200$ saved snapshots.\footnote{This package does not natively support perturbations, so for the perturbed cases we use a slightly hand-modified version of the original code.}

\smallskip

\begin{figure}[htb]
    \centering
    \includegraphics[width=0.995\linewidth]{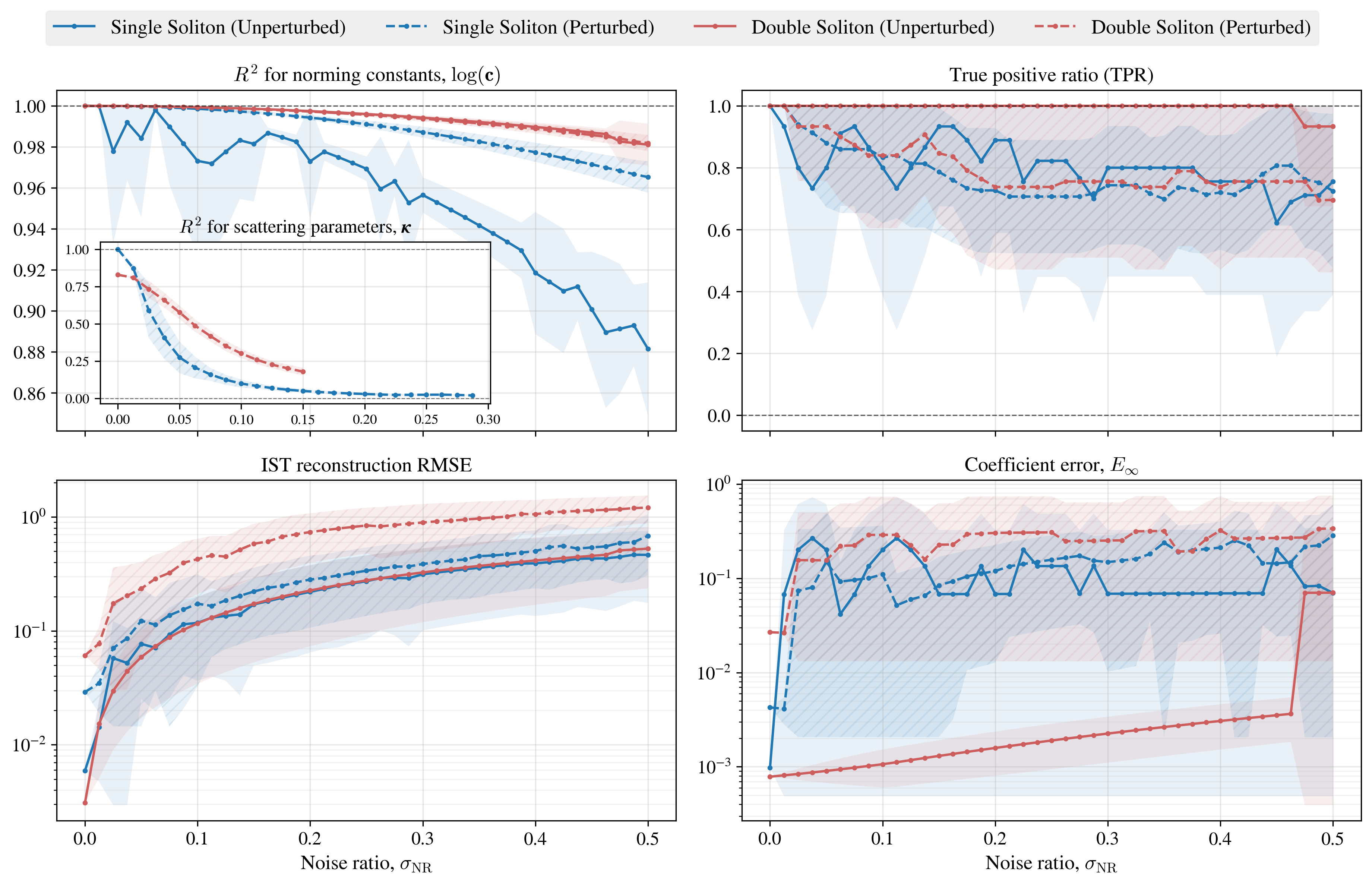}
    \caption{Illustrating how the validation metrics of \S\ref{subsec:validation} scale with increasing noise for each of the four configurations of synthetic data. The noise ratio $\sigma_{\textsc{nr}}$ is swept from $0$ to $0.5$ in increments of $\Delta\sigma_{\textsc{nr}} = 0.0125$ using $15$ distinct realizations per nonzero noise level; solid lines denote sample means while the shaded bands denote $\pm 1$ standard deviation. [Top-left] $R^2$ values for the identified norming constant models, with an inset corresponding to the scattering parameter models; as per \S\ref{subsec:validation}, we omit $R^2$ values for runs in which the zero model $\hat{\mathbf{w}} = 0$ was selected. The TPR [top-right], coefficient error [bottom-right], and IST reconstruction error [bottom-left] are computed via eqns.~(\ref{eq:Einfty_and_TPR}) and (\ref{eq:IST_reconstruction_RMSE})}
    \label{fig:validation_plots}
    \vspace{-2mm}
\end{figure}

In the unperturbed case, the scattering parameters $\kappa_i$ are conserved quantities (i.e., $\dot{\kappa}_i = 0$), meaning that any single simulation samples only a constant value for each parameter. Consequently, the candidate monomials $\kappa_i,\, \kappa_i^2,\, \kappa_i^3$ in the WSINDy library of eq.~(\ref{eq:candidate_library}) are linearly dependent along the trajectory, and the regression problem in eq.~(\ref{eq:discrete_weak_form_loss}) becomes ill-posed. To account for this degeneracy, we stack scattering data $\mathbf{X}_r$ from three distinct simulations $r = 1,2,3$ (the minimum number required for well-posedness) and concatenate each respective weak-form linear system from eq.~(\ref{eq:weak_linear_system}) into a single block-regression. This is motivated by the nature of the experimental dataset of \cite{Heinrich.etal2025} investigated in \S\ref{subsec:empirical_data} below, which includes several (25) distinct experimental runs. Each simulation is initialized using sech-squared profiles $u_0(x) = \sum_{i=1}^{n} 2\kappa^2_i\text{sech}^2(\kappa_i(x - x_i))$ parameterized by a distinct set of scattering parameters $\kappa_i$ and initial positions $x_j := 5(j - 5)$. For the single soliton examples in which $n=1$, we vary $\kappa \in \{2, 2.2, 2.4\},$ while for the two-soliton cases with $n=2$, we instead use $(\kappa_1,\kappa_2) \in \{(2, 1.8), (2.2, 1.4), (2.4, 1.2)\}$; the $\kappa = 2$ and $(\kappa_1,\kappa_2) = (2.4, 1.2)$ cases are shown in Figure~\ref{fig:kdv_bright_soliton_snapshots}. Finally, because the unperturbed dynamics obey a trivial conservation law of the form $\dot{\boldsymbol{\kappa}} = 0$, we extend the standard MSTLS of Algorithm~\ref{algorthim:MSTLS} to the zero-admissible formulation given in Algorithm~\ref{algorthim:BIC_MSTLS}, which uses BIC-based hypothesis testing to allow WSINDy to select model coefficients with empty support.

\smallskip

As reported in Table~\ref{table:kdv-synthetic-results}, the recovered coefficients agree with both the exact scattering ODEs and the leading-order perturbation theory to within a fraction of a percent at zero noise; e.g., in the perturbed single soliton example, WSINDy estimates $\dot{\kappa} \approx 0.0667\kappa$ versus the ground-truth of $\dot{\kappa} = \tfrac{1}{3} \epsilon\kappa \approx 0.0666\kappa$. In turn, Figure~\ref{fig:validation_plots} tracks how the model identification results deform as the noise ratio $\sigma_{\textsc{nr}}$ increases, revealing a few consistent trends. In particular, the weak-form regression for the $\log(\mathbf{c})$ models stays robust across the entire noise range -- the $R^2$ for these ODEs remains above $\sim\!0.96$ for three of the four configurations and declines only mildly to $\sim\!0.88$ for the weakest (i.e., unperturbed single-soliton) case at $50\%$ noise. By contrast, the $R^2$ values decay rapidly for the perturbed $\dot{\boldsymbol{\kappa}}$ equations, which feature small $\mathcal{O}(\epsilon)$ coefficients. Mechanistically, beyond $\sim\!15\%$ background noise this perturbation becomes indistinguishable from a conservation law of the form $\dot{\kappa} = 0$, at which point the BIC test in Algorithm~\ref{algorthim:BIC_MSTLS} selects the zero model $\hat{\mathbf{w}} = 0$ and a corresponding $R^2$ value is no longer reported.

\newpage

Although the $R^2$ eventually becomes unstable for the perturbed cases featuring small coefficients, the correct model terms are nonetheless reliably recovered: the mean TPR stays above $\sim\!0.7$ at all noise levels for each of the four configurations, while the unperturbed two-soliton example in particular retains a TPR of exactly one until $\sim\!45\%$ noise. In a similar vein, the $E_{\infty}$ coefficient error remains $\mathcal{O}(\texttt{1e-3})$ across nearly the entire noise range in the unperturbed two-soliton case, rising only beyond $\sigma_{\textsc{nr}} \approx 0.45$. By contrast, in the single-soliton and perturbed cases, the $E_{\infty}$ coefficient error plateaus near $\mathcal{O}(\texttt{1e-1})$, again reflecting the difficulty of resolving the small perturbative coefficients. Moreover, occasional sharp oscillatory peaks are present in several cases -- features which we predominantly attribute to sampling error. Lastly, we observe that the reconstruction RMSE, a proxy for the predictive fidelity of the identified models, grows monotonically with $\sigma_{\textsc{nr}}$ in every case. Unsurprisingly, the most dynamically complex configuration (i.e., the perturbed two-soliton collision) tends to incur the largest reconstruction error.

\vspace{-2.5 mm}

\subsection{Empirical Data}\label{subsec:empirical_data}
\begin{figure}[htb]
    \centering
    \raisebox{7mm}{\includegraphics[width=0.56\linewidth]{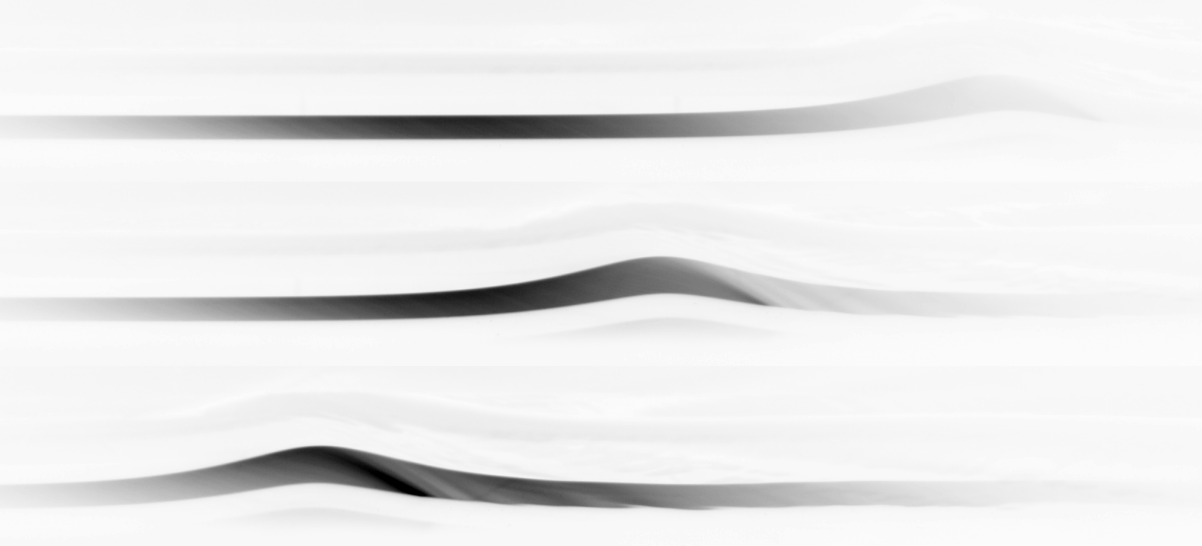}}
    \includegraphics[width=0.43\linewidth]{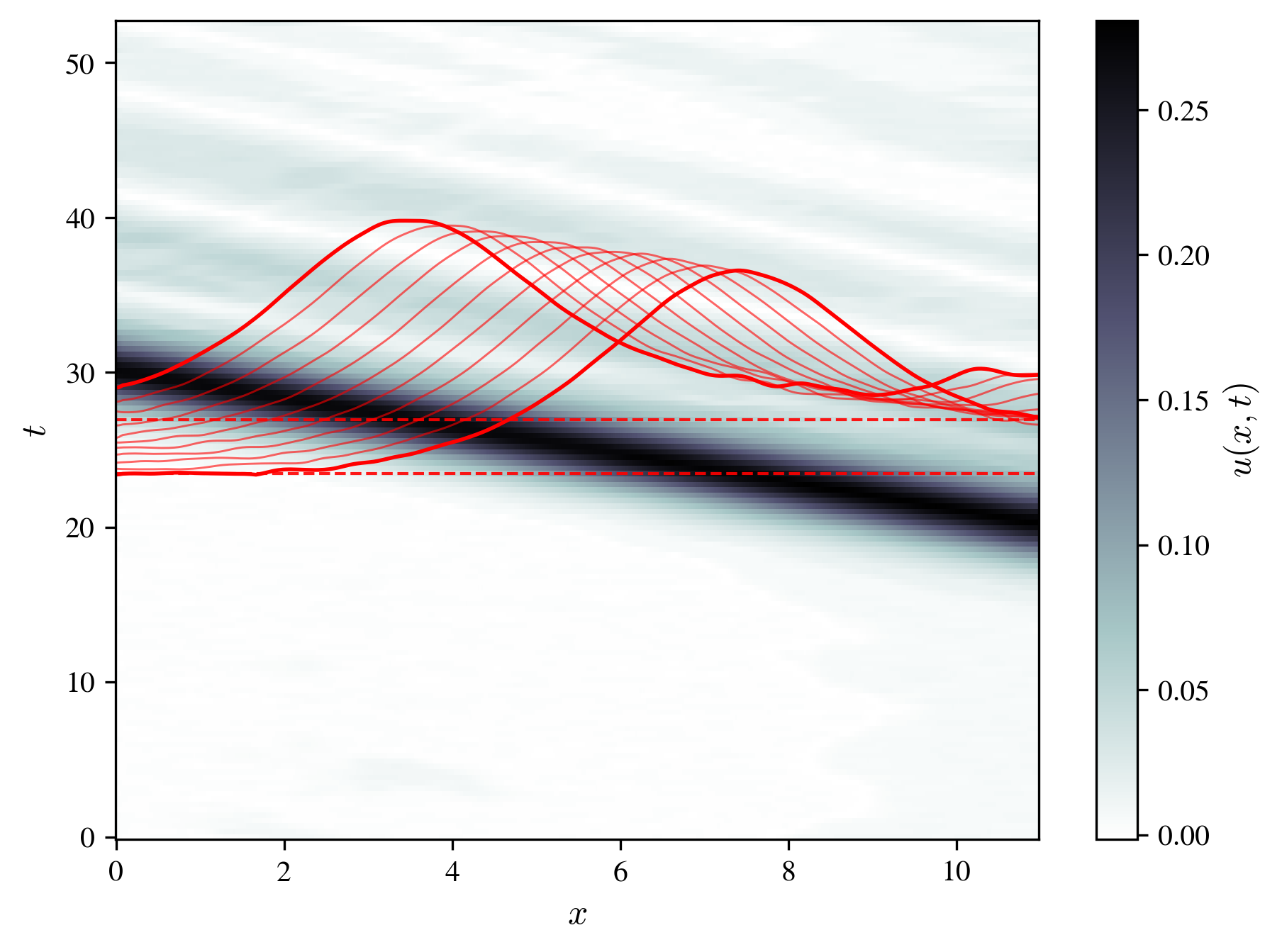}
    \includegraphics[width=\linewidth]{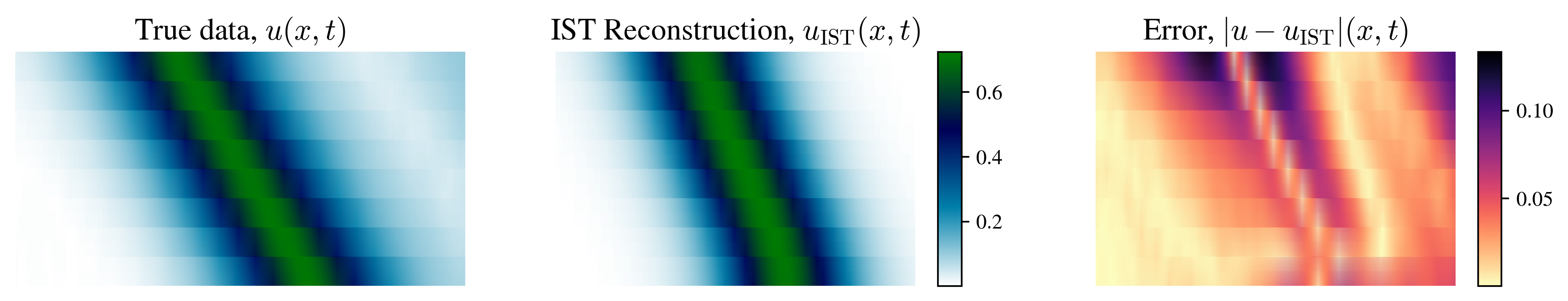}
    \caption{[Top-left] Example snapshots from the experimental dataset of \cite{Heinrich.etal2025, Heinrich.etal2026}. [Top-right] The selected temporal window for one trial in which the soliton sits solidly in-frame, showing the extracted profiles (red) overlaid on the measured field $u = u(x,t)$. [Bottom] Visualizing, for one held-out validation trial, the measured field, the reconstruction obtained by forward-integrating the WSINDy-discovered scattering dynamics and applying the IST, and the corresponding pointwise error}
    \label{fig:heinrich_et_al_data}
\end{figure}

In this section, we illustrate the discovery of effective soliton dynamics using real experimental data. Specifically, we consider data from a recent study by Heinrich et al. \cite{Heinrich.etal2025, Heinrich.etal2026} in which a Kawahara-type PDE was identified from video recordings of solitary shallow-water waves propagating down a flume apparatus (see Figure~\ref{fig:heinrich_et_al_data}). In this study, the edge-detection algorithm of \cite{Canny1986IEEETransPatternAnalMachIntell} was used to extract a nondimensionalized\footnote{The rescaling is $u^*\!,x^*\!,t^* \mapsto hu, hx, \sqrt{h/g}t$, where $u^*\!,x^*\!,t^*$ are the measured dimensional quantities and $h \approx 32$ mm is the water depth.}  surface field $u_k(x,t)$ for each of $k = 1, \dots, 25$ distinct experimental trials, $18$ of which were subsequently concatenated into a single data-matrix $\mathbf{U}$ (cf. \S\ref{subsec:synthetic_data} above). The authors then applied two distinct equation-learning methodologies to the data -- WSINDy and a novel "Fourier multiplier" method \cite{Heinrich.etal2025} -- both of which independently selected a sparse model of the form \begin{equation}\label{eq:heinrich-pde}
  u_t + v_0u_x + \alpha uu_x + \beta u_{xxx} = \epsilon u_{xxxxx}.
\end{equation} Moreover, similar coefficient values were estimated by each method (e.g., $\epsilon \approx -0.06$ from both methods). In a moving frame of reference $\xi = x - v_0t$, this model is equivalent to the Kawahara model in eq.~(\ref{eq:kawahara_eqn}), \begin{align*}
    u_t + \alpha uu_{\xi} + \beta u_{\xi\xi\xi} = \epsilon u_{\xi\xi\xi\xi\xi}.
\end{align*} To validate the discovered model, the authors of \cite{Heinrich.etal2025} numerically integrated eq.~(\ref{eq:heinrich-pde}) forward in time and compared the predicted field to the measured soliton surface heights from each of the seven held-out experimental trials, achieving errors of just $4$-$6\%$ relative to the wave amplitudes.

\smallskip

Note that, for our purposes here, we apply two additional lightweight preprocessing steps to the data: (1) a small number of spike artifacts were repaired via local temporal averaging; (2) we subtracted a small offset from the water level $u \mapsto u - \bar{u}$, where $\bar{u} \sim 0.02$--$0.13$ mm, so that $u \rightarrow 0$ as $x \rightarrow \pt\!\Omega$. Just as in \cite{Heinrich.etal2026}, each run is sampled over $N_x = 1200$ points in space ($\Delta{x} \approx 0.29$ mm) and $N_t = 151$ frames in time ($\Delta{t} = 0.02$ s). In each case, we isolate a specific temporal window in which the soliton stays solidly in-frame, which we then temporally super-sample via cubic-spline interpolation before computing the scattering data (see Figure~\ref{fig:heinrich_et_al_data}). We estimate that the experimental noise level is roughly $\sigma_{\textsc{nr}} \approx 1.4\%$, comfortably within the regime in which WSINDy provides reliable model estimates (cf.~\S\ref{subsec:synthetic_data}).


\vspace{-1.5 mm}
\begin{table}[htbp]
\TBL{
\caption{The models identified from the experimental data of \cite{Heinrich.etal2025, Heinrich.etal2026}. Here, we report the RMSE of eq.~(\ref{eq:IST_reconstruction_RMSE}) normalized to the peak amplitude $A := \max|u|$ and averaged over the seven validation trials}
\label{table:empirical-results}
}
{%
\small
\renewcommand{\arraystretch}{1.55}

\begin{tabular*}{\linewidth}{@{\extracolsep{\fill}}%
L{0.4\linewidth}
L{0.3\linewidth}
L{0.14\linewidth}
L{0.14\linewidth}@{}}

\toprule
\TCH{Identified model} &
\TCH{Library} &
\TCH{RMSE} &
\TCH{$R^2$}
\\
\midrule

\smallresultrow
{\resultcell{
    \dot{\kappa} = {\color{purple}0},
    \\
    \tfrac{d}{dt}\log(c) = {\color{purple}3.46}\kappa + {\color{purple}2.60}\kappa^3 + {\color{purple}0.20}\log(c)
}}
{\resultcell{
    \{1\} \cup \{\kappa^p, \log(c)^p\}_{1 \leq p \leq 3},
    \\
    \{1, \kappa, \kappa^3\} \cup \{\log(c)^p\}_{1 \leq p \leq 3},
}}
{17.6\%}
{0.803}

\botrule
\end{tabular*}
}
\end{table}
\clearpage


\begin{figure}[htb]
    \centering
    \includegraphics[width=0.92\linewidth]{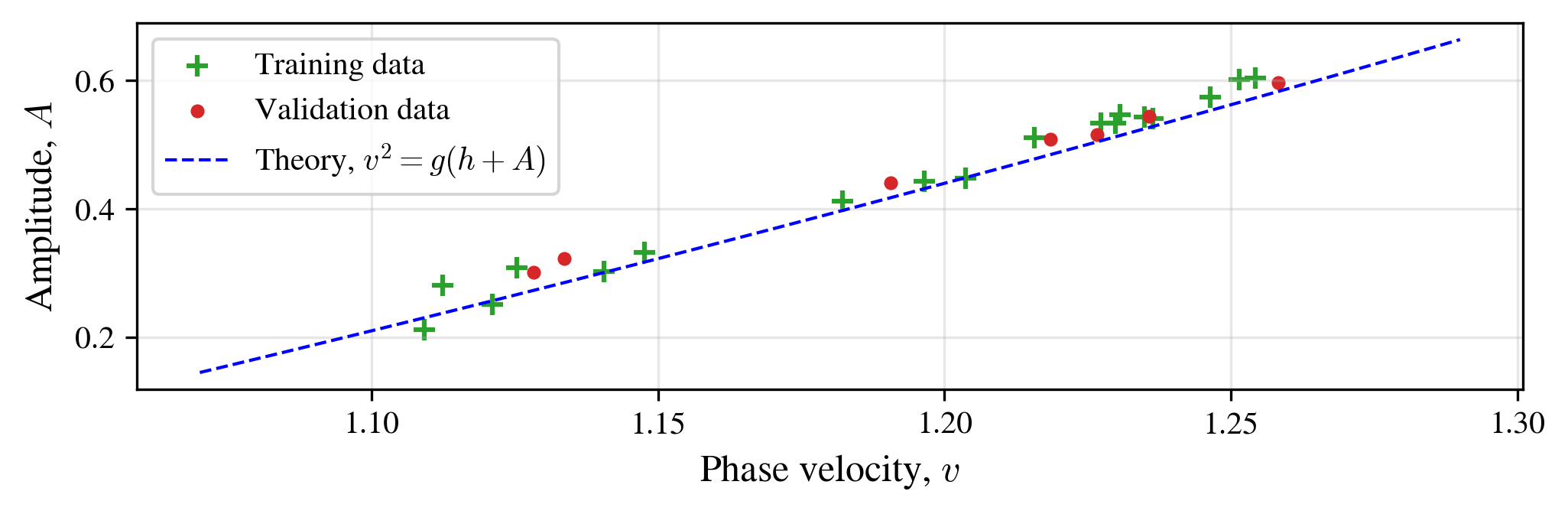}
    \caption{A reproduction of Figure~9 from Ref.~\cite{Heinrich.etal2025}, which illustrates an empirical amplitude--phase velocity relation obeyed by the shallow-water waves. Here, the dimensional amplitudes $A^* := \max|hu^*|$ are plotted against the phase-velocity $v^* := |\dot{x}|/\sqrt{gh}$ for the $18$ training and $7$ validation trials, showing the validity of the empirical relation. Note that the first two terms in the identified $\tfrac{d}{dt}\log(c)$ model listed in Table~\ref{table:empirical-results} are consistent with first two terms of a Taylor series expansion of the nondimensionalized $v \approx \sqrt{1+A}$ relationship; see \S\ref{subsec:empirical_data}}
    \label{fig:phasevelocity_vs_amplitude}
\end{figure}

For each equation listed in Table~\ref{table:empirical-results}, we use a candidate library inspired both by the identified PDE model in eq.~(\ref{eq:heinrich-pde}) and by the empirical amplitude--phase-velocity relation $v^2 \approx 1 + A$ reported in \cite{Heinrich.etal2025},\footnote{See Figure~9 therein.} where $A := \max|u(x,t)|$ and $v := |\dot{x}|$ respectively denote the nondimensionalized amplitude and phase velocity of the traveling water wave (see Figure~\ref{fig:phasevelocity_vs_amplitude}). In particular, a single exact KdV soliton satisfies $\log(c) = 2\kappa\xi + \log(2\kappa)$ in a co-moving frame of reference $\xi(t) = \xi_0 + 4\beta\kappa^2t$ (cf. eq.~(\ref{eq:scattering_ODEs})); however, a stationary observer (i.e., in the lab frame $x = \xi + v_0t$) would measure an additional advective force, \begin{align}\label{eq:theory_ODE}
    \tfrac{d}{dt}\log(c) = 2\kappa{v} = 2v_0\kappa + 8\beta\kappa^3.
\end{align} Inserting an expansion of the empirical relation $v \approx \sqrt{1 + A} = 1 + (1/2)A + \mathcal{O}(A^2)$ into eq.~(\ref{eq:theory_ODE}) and using the exact soliton amplitude $A(\kappa) = -(2/\gamma)\kappa^2 = 12(\beta/\alpha)\kappa^2$ independently yields \begin{align}\label{eq:empirical_ODE}
    \tfrac{d}{dt}\log(c) = 2v\kappa
    \approx 2\kappa\left(1 + \tfrac{6\beta}{\alpha}\kappa^2 + \mathcal{O}(\kappa^4)\right)
    = 2\kappa + \tfrac{12\beta}{\alpha}\kappa^3 + \mathcal{O}\big(\kappa^5\big).
\end{align} Consistency between the idealized and empirical ODEs in eqs.~(\ref{eq:theory_ODE}) and (\ref{eq:empirical_ODE}) requires that both $v_0 = 1$ (empirically, $v_0 \approx 1.2$) and $8\beta = 12(\beta/\alpha)$, which occurs if and only if $\alpha = 3/2$ ($\alpha \approx 1.4$ and $8\beta \approx 5$). Therefore, the observed $v^2 \approx 1 + A$ relationship may be viewed as a restatement of the ODE in eq.~(\ref{eq:empirical_ODE}). Guided by this structure, we use a candidate library based on the odd powers $\{\kappa, \kappa^3\}$, where we interpret the linear term as capturing advection and the cubic term as capturing the soliton's nonlinear evolution, together with low-order powers of $\log(c)$ (see Table~\ref{table:empirical-results}). Although the identified Kawahara-type PDE is translation-invariant, our hope is that the latter terms allow us to absorb any residual position-dependence introduced by the experimental or numerical setup rather than by the physical dynamics.\footnote{Recall that $\log(c)$ is highly correlated with the soliton position $\xi$.}


\clearpage

\vspace{-3 mm}

Using the setup described above, WSINDy selects the conservation law $\dot{\kappa} = 0$ together with a model for $\tfrac{d}{dt}\log(c)$ that is consistent with the form of the cubic expansion given in eq.~(\ref{eq:theory_ODE}), with the exception of an additional small $\log(c)$ term reminiscent of the perturbed scattering ODEs from Table~\ref{table:kdv-synthetic-results} above. We emphasize, however, that we do not realistically expect the additional $\log(c)$ term to play an active role in the physical dynamics -- instead, we attribute its existence to experimental and numerical artifacts, and note that it likely skews the relative magnitudes of the identified coefficients (cf. eq.~(\ref{eq:theory_ODE})). Forward integration of the discovered scattering dynamics and subsequent reconstruction of the water-line via the IST yields an amplitude-normalized RMSE of $8$--$27\%$ (or $\sim\!18\%$ on average; see Table~\ref{table:empirical-results}) across the seven held-out trials (cf. Figure~\ref{fig:heinrich_et_al_data}). In each case exhibiting large RMSE, the reconstruction errors are predominantly caused by a secondary wave trailing behind the main soliton (i.e., additional radiation), which the reflectionless IST is not capable of representing.

\vspace{-2 mm}

\subsection{A Note on Identifiability}\label{subsec:identifiability}
We conclude the results section with a few brief remarks on the identifiability of nonlinear dynamics from solitary wave trajectories. Rudy et al. \cite{Rudy.etal2017SciAdv} observed that when SINDy-based methods are applied to measurements of a single soliton evolving along a single trajectory, they tend to recover trivial linear transport equations in place of any underlying nonlinear wave dynamics.\footnote{As in \S\ref{subsec:synthetic_data}, this problem can be avoided by stacking trajectories -- for example, Heinrich et al. \cite{Heinrich.etal2025} use 18 distinct single soliton trajectories.} Interestingly, this behavior represents an inherent feature of the data, rather than a deficiency of the SINDy architecture; i.e., every lone KdV soliton \textit{exactly} solves a corresponding linear transport equation, \begin{align}\label{eq:transport_eqn}
    u(x,t) = 2\kappa^2\,\text{sech}^2\!\big(\kappa\big(x - x_0 - 4\kappa^2 t\big)\big)
    \quad \text{solves both:} \quad
    \begin{cases}
        u_t + 4\kappa^2 u_x = 0,
        \\
        u_t + 6uu_x + u_{xxx} = 0.
    \end{cases}
\end{align} Since both candidate models in eq.~(\ref{eq:transport_eqn}) reproduce the data exactly, a sparse-regression architecture naturally selects the simpler one -- in this case, the transport equation is a more parsimonious model than the KdV equation. Physically speaking, nonlinearity manifests itself in the KdV equation via the dependence of wave-speeds on wave-amplitudes, and no such dependence can be observed from a soliton in isolation. Conversely, nonlinearities can be identified by, e.g., observing a two-soliton interaction or, more generally, a collection of solitons of differing amplitudes \cite{Rudy.etal2017SciAdv}. A closely-related identifiability issue was subsequently reported by Vasey et al. \cite{Vasey.etal2025JournalofComputationalPhysics} in the setting of magnetohydrodynamic blast waves.

\newpage

Working with scattering data $(\kappa_i, c_i)(t)$ instead of field data $u(x,t)$ does not fundamentally remove the degeneracy described above, although it does change the nature of the identifiability problem in an interesting and potentially useful way -- in particular, the scattering ODEs in eqs.~(\ref{eq:scattering_ODEs}) and (\ref{eq:scattering_ODEs_perturbed}) are nonlinear in $\kappa_i$, regardless of the number of solitons present in the data. However, in analogy with eq.~(\ref{eq:transport_eqn}), each $\kappa_i(t) = \kappa_i(0)$ is conserved and thus constant in the absence of an external perturbation; consequently, the candidate monomials $\{\kappa_i,\,\kappa^2_i,\,\kappa^3_i\}$ in the library of eq.~(\ref{eq:candidate_library}) are nearly collinear along a single trajectory, and linear regression cannot uniquely determine which monomial governs the evolution of $\log(c_i)$. It is important to recognize that the collinearity of the $\kappa^p_i$ is the manifestation, observed at the level of the scattering data, of the very same mechanism responsible for the spurious transport equation observed at the PDE level. That being said, the nature of this degeneracy does become more benign in the scattering coordinates -- the recovered model never collapses onto a structurally distinct equation. The conservation laws $\dot{\kappa}_i = 0$ and the linear-in-time growth of $\log(c_i)$ can still be identified, and only the exponents of the $\tfrac{d}{dt}\log(c_i) = \kappa^p_i$ models remain unknown. As noted above, the identifiability problem dissolves as the data either begin to feature either more than one soliton or more than one trajectory, the latter of which is a phenomenon that has been observed in settings as diverse as network dynamics \cite{Tian.etal2026}.

\smallskip

To concretely illustrate the above points, we observe that applying WSINDy for PDEs \cite{Messenger.Bortz2021JournalofComputationalPhysics} to just the first experimental training dataset from \cite{Heinrich.etal2026} yields the transport equation $u_t \approx 1.2u_x$ with $R^2 \approx 0.99$. Interestingly, when using the scale-invariant preconditioning method described in \cite{Messenger.etal2024SciRep}, WSINDy instead identifies a nonlinear PDE of the form $u_{t} \approx - 0.0u + 0.1u^2 + 0.9u_{x} + 0.8(u^2)_{x}$ where again $R^2 \approx 0.99$, indicating that scaling becomes particularly important when attempting to recover such nonlinearities.\footnote{We note that additional radiation content in the experimental data may also have aided in the identification of the nonlinear $(u^2)_x$ term.} In a similar vein, regressing the scattering data against a simple polynomial library of the form $\{\kappa,\kappa^2,\kappa^3\}$ (evaluated over the same experimental trial data) yields an advective law of the form $\tfrac{d}{dt}\log(c) \approx -2\kappa,$ irrespective of whether the library columns are rescaled. In this case, $\kappa(t) \approx 2/3$ is nearly constant along the trajectory, and the identified model is almost indistinguishable from both $-3\kappa^2$ and $-5\kappa^3$.

\vspace{-2.95 mm}

\section{Discussion}\label{sec:discussion}
\vspace{-1.95 mm}
In this paper, we have proposed and investigated a data-driven method for modeling effective soliton dynamics that works directly with scattering data. In particular, this technique combines the conceptual framework of the IST \cite{Ablowitz.etal1974StudApplMath} with the weak-form equation-learning paradigm of WSINDy \cite{Messenger.Bortz2021MultiscaleModelSimul} to identify interpretable symbolic models without requiring prior knowledge of the form of the scattering equations. Parameterizing a scalar field $u$ via its scattering data $(\kappa_i, c_i)$ is natural in the context of near-integrable wave equations and inherits a large body of analytical structure from existing theory, e.g., the fact that unperturbed KdV scattering dynamics take a convenient polynomial form in $(\kappa_i, \log(c_i))$ coordinates. We have focused on shallow-water models of KdV-type and demonstrated that for this class of governing equations, the technique reliably recovers both the exact unperturbed scattering ODEs as well as the leading-order corrections predicted by the perturbation theory \cite{Karpman.Solovev1981PhysicaDNonlinearPhenomena}.

\smallskip

Our numerical experiments indicate that the inferred scattering dynamics are robust to low amounts of additive i.i.d. Gaussian measurement noise, both in unperturbed and perturbed regimes. On synthetic data, the recovered coefficients agree with the exact and perturbed scattering ODEs to within a fraction of a percent at $0\%$ noise, and the $\tfrac{d}{dt}\log(\mathbf{c})$ models remain accurate (i.e., $R^2 \gtrsim 0.96$ in three of four cases)\\ across the entire $0$--$50\%$ noise range tested. Unsurprisingly, the small $\mathcal{O}(\epsilon)$ corrections to the $\dot{\boldsymbol{\kappa}}$ equations are harder to resolve in the presence of measurement noise; beyond $\sigma_{\textsc{nr}} \gtrsim 0.15$, these data become statistically indistinguishable from the conservation law $\dot{\boldsymbol{\kappa}} = 0$ and the BIC test of Algorithm~\ref{algorthim:BIC_MSTLS} defaults to the zero model. When applied to the experimental shallow-water wave dataset of Heinrich et al. \cite{Heinrich.etal2025, Heinrich.etal2026}, the method recovers $\dot{\kappa} = 0$ together with a $\tfrac{d}{dt}\log(c)$ model consistent with a cubic expansion of the empirical amplitude--phase-velocity relationship reported therein; forward-integration of the recovered dynamics and a subsequent IST reconstruction $(\kappa,\log(c)) \mapsto u$ yields an amplitude-normalized RMSE of $\sim\!18\%$, with the largest errors stemming from the existence of excess radiation trailing the main wave.

\clearpage

Notably, numerical experiments that featured a single, unperturbed soliton consistently saw worse performance than their two-soliton counterparts across each validation metric ($R^2$, $E_\infty$, RMSE, TPR), despite the fact that these cases are significantly simpler from a dynamical perspective. We attribute this somewhat counterintuitive result to a structural identifiability mechanism similar to that observed in both the KdV and magnetohydrodynamic settings of \cite{Rudy.etal2017SciAdv} and \cite{Vasey.etal2025JournalofComputationalPhysics}, in which solitary waves measured along a single trajectory were found to obey advection equations that were much simpler than the expected underlying nonlinear dynamics. Although working with scattering data does not fundamentally remove this degeneracy, it does change the nature of the identifiability problem in a convenient way -- while lone solitons exactly obey linear dynamics at the PDE level, the ODEs governing their scattering data are, by contrast, inherently nonlinear in $\kappa_i$ regardless of the number of solitons present. In the scattering formulation, this degeneracy enters only at a more benign numerical conditioning level, with constant $\kappa_i$ values rendering the candidate monomials $\kappa^p_i$ collinear. Moreover, this ill-conditioning is avoided in perturbed regimes where the $\kappa_i$ are non-constant, and in these cases the identified $\mathcal{O}(\epsilon)$ corrections conceivably allow one to back out a set of consistent forms for the forcing term $F[u]$.

\smallskip

Although we have focused here on shallow-water equations of KdV-type, a similar pipeline applies, in principle, to any PDE admitting a Lax-pair representation (e.g., any integrable wave equation from \cite{Ablowitz.etal1974StudApplMath}).\\ Practically speaking, extending the framework discussed here to a new integrable system would require three main ingredients: \begin{enumerate}
    \item the Lax operator corresponding to the PDE in question;

    \smallskip
    
    \item a numerical DST scheme for the associated EVP;

    \smallskip
    
    \item a choice of coordinates in which the scattering dynamics can be reasonably well-approximated by a low-order library of candidate terms.
\end{enumerate} Applying this pipeline to the focusing nonlinear Schr\"odinger equation would be a natural next step, subject to the caveat that a treatment of this equation would introduce additional difficulties that haven't been considered herein. In particular, rather than the self-adjoint EVP considered above, its scattering data are defined in terms of a non-self-adjoint EVP (i.e., the $2\!\times\!2$ Zakharov-Shabat system) that has a fundamentally complex-valued point spectrum featuring non-trivial real \textit{and} imaginary components. Additionally, addressing the defocusing regime (in which dark solitons sit atop a nonzero background), would require developing a strategy for relaxing the assumption of decay as $x \rightarrow \infty$.


\smallskip

We conclude by considering a few natural extensions of the current work. In addition to addressing other integrable PDEs and more straightforward methodological improvements such as an end-to-end data-driven pipeline utilizing WSINDy \cite{Messenger.Bortz2021JournalofComputationalPhysics} and SILO \cite{Adriazola.etal2026SIAMJApplDynSyst}, one could consider accommodating nonzero radiation (i.e., $R \neq 0$) by appending reflection coefficients to the state vector as outlined in \S\ref{subsec:outline}. This would generalize the modeling technique to non-solitonic settings and, seeing as how excess radiation was the dominant source of reconstruction error in our experimental results, would likely increase the predictive capacity of the discovered models. From a more theoretical perspective, the inclusion of radiation might also lead one to consider \textit{purely dispersive} regimes in which $\mathbf{c} = 0$ while $R \neq 0$, including connecting the long-time asymptotics of such systems to Painlev\'e-type equations \cite{ablowitzSolitonsNonlinearEvolution1991}. Lastly, we expect that it would be interesting and potentially fruitful to explore geophysical modeling applications related to, e.g., internal ocean waves, coastal shoaling waves, and morning-glory waves -- settings where coherent nonlinear wave phenomena are ubiquitous and high-resolution field data are becoming increasingly available \cite{elDispersiveShockWaves2016, Lee.etal2024PLoSONE}.

\newpage

\begin{appendix}

{\color{black}
\begin{algorithm}
    {\color{black}
    \caption{Modified Sequential Thresholding Least Squares ($\texttt{MSTLS}$)}\label{algorthim:MSTLS}
    \textbf{Inputs:} response vector $\mathbf{b}$ (size $K$), library matrix $\mathbf{G}$ (size $K \times J$), thresholding parameter $\lambda \in (0,1)$.

    \textbf{Outputs:} thresholded weights $\mathbf{w}^{\lambda}$ (size $J$).
    
    \vspace{-2mm} \hrulefill \vspace{-2mm}
    \begin{enumerate}
        \item $n \leftarrow 0$
        
        \item $\mathbf{w}^0 \leftarrow \mathbf{w}_{\textsc{ls}}$

        \item $\texttt{stopping\_criterion} \leftarrow \texttt{false}$

        \item \textbf{for} $j = 1, \dots, J$ \textbf{do}: \begin{itemize}
            \item $L_j \leftarrow \max\big(1, \, \|\mathbf{b}\|_2/\| \mathbf{G}_j \|_2\big)$

            \item $U_j \leftarrow \min\big(1, \, \|\mathbf{b}\|_2/\| \mathbf{G}_j \|_2\big)$
        \end{itemize}

        \item \textbf{while} \texttt{stopping\_criterion == false} \textbf{do}: \begin{itemize}
            \item $\mathcal{I}_{n} \leftarrow \big\{ 1 \leq j \leq J \, : \, |\mathbf{w}^n_j| \in [\lambda L_j, \, \lambda^{-1} U_j] \big\}$
            
            \item $\mathbf{w}_{n+1} \, \leftarrow \ \text{arg}\!\min_{\text{supp}(\mathbf{w}) \subseteq \mathcal{I}_n} \|\mathbf{b} - \mathbf{Gw}\|^2_2$

            \item \textbf{if} $n \geq 1$ \textbf{and} $\mathcal{I}_{n} = \mathcal{I}_{n-1}$: \begin{itemize}
                \item[$\circ$] $\texttt{stopping\_criterion} \leftarrow \texttt{true}$

                \item[$\circ$] $\mathbf{w}^{\lambda} \leftarrow \mathbf{w}_{n+1}$
            \end{itemize}

            \item $n \leftarrow n+1$
        \end{itemize}
    \end{enumerate}}
\end{algorithm}}

\vspace{-7mm}

\section{The MSTLS Algorithm}\label{appendixA:MSTLS}
We use a zero-admissible extension of the Modified Sequential Thresholding Least Squares (MSTLS) algorithm developed in \cite{Messenger.Bortz2021JournalofComputationalPhysics} to approximately solve the sparse regression problem posed in eq.~(\ref{eq:sparse_regression_problem}). {\color{black}In MSTLS, a sparse vector of model weights $\hat{\mathbf{w}}_i \in \mathbb{R}^J$ with $\|\hat{\mathbf{w}}_i\|_0 \ll J$ for each $i=1,\dots,d$ is obtained by minimizing a normalized version of the loss function given in eq.~(\ref{eq:sparse_regression_problem}) over a finite set of \textit{thresholding parameters} $\boldsymbol{\lambda} := \{\lambda_l : l = 1, \dots, N_{\lambda}\} \subset (0,1)$.} In particular, the model weights are explicitly given by \begin{align}\label{eq:wsindy_loss}
    \hat{\mathbf{w}}^{\textsc{mstls}}_i := \texttt{MSTLS}\big(\mathbf{b}_i, \, \mathbf{G}, \, \lambda^{\star}\big),
    \quad \text{with} \quad
    \lambda^* := \min \left[\argmin_{\lambda \in \boldsymbol{\lambda}} \, \mathcal{L}_{\textsc{mstls}}(\lambda) \right],
\end{align} where $\texttt{MSTLS}$ denotes the output of Algorithm~\ref{algorthim:MSTLS}. For a given thresholding parameter $\lambda \in (0,1)$, the loss function $\mathcal{L}_{\textsc{mstls}}$ is defined as \begin{align*}
    \mathcal{L}_{\textsc{mstls}}(\lambda) := \mathcal{L}_{\mu}\!\left(\mathbf{w}^{\lambda}_i; \, \frac{\mathbf{b}_{i}^{\textsc{ls}}}{\| \mathbf{b}_{i}^{\textsc{ls}} \|_2}, \, \frac{\mathbf{G}}{\| \mathbf{b}_{i}^{\textsc{ls}} \|_2}\right),
    \quad \text{with} \quad
    \mu := \frac{1}{J},
\end{align*} where $\mathbf{b}^{\textsc{ls}}_{i} := \mathbf{G}\mathbf{w}^{\textsc{ls}}_{i}$ is the projection of the ordinary least-squares estimate given by \begin{align*}
    \mathbf{w}^{\textsc{ls}}_{i} := \big(\mathbf{G}^T\mathbf{G}\big)^{-1}\mathbf{G}^T\mathbf{b}_i.
\end{align*} Here, the quantity $\mathbf{w}^{\lambda}_i := \texttt{MSTLS}(\mathbf{b}_i, \mathbf{G}, \lambda)$ denotes the vector of $\lambda$-thresholded weights, which at every iteration satisfies the following dominant balance relationship: \begin{align*}
    \frac{\|w^{\lambda}_{\!ji}\mathbf{G}_j\|_2}{\|\mathbf{b}_i\|_2} \in \big[\lambda, \, \lambda^{-1}\big],
    \quad \text{for each} \quad
    j = 1, \dots, J.
\end{align*} We follow \cite{Messenger.Bortz2021JournalofComputationalPhysics} in scanning over a set of $N_{\lambda} = 50$ candidate values $\boldsymbol{\lambda} := \{\lambda_l\}^{50}_{l=1}$ defined by uniformly log-spaced increments $\log_{10}(\lambda_l) \in (-4,0)$ with the exception of the perturbed $\log(c)$ cases in \S\ref{subsec:synthetic_data} and \S\ref{subsec:empirical_data}, for which we found it helpful to hand-tune the $\lambda^*$ parameter to $\texttt{4e-2}$ and $\texttt{1e-1}$, respectively.

\clearpage

{\color{black}
\begin{algorithm}
    {\color{black}
    \caption{MSTLS with BIC Testing for Trivial Dynamics}\label{algorthim:BIC_MSTLS}
    \textbf{Inputs:} response vector $\mathbf{b}$ (size $K$), library matrix $\mathbf{G}$ (size $K \times J$), thresholding parameters $\boldsymbol{\lambda} = \{\lambda_l\}_{l=1}^{N_{\lambda}}$.

    \textbf{Outputs:} BIC-tested weights $\hat{\mathbf{w}}$ (size $J$).
    
    \vspace{-2mm} \hrulefill \vspace{-2mm}
    \begin{enumerate}
        \smallskip
        
        \item $\lambda^* \leftarrow \min \left[\argmin_{\lambda \in \boldsymbol{\lambda}} \, \mathcal{L}_{\textsc{mstls}}(\lambda) \right]$

        \smallskip
        
        \item $\hat{\mathbf{w}}^{\textsc{mstls}} \leftarrow \texttt{MSTLS}\big(\mathbf{b}, \, \mathbf{G}, \, \lambda^*\big)$

        \smallskip

        \item $\Delta \leftarrow \gamma^2\log\left(\|\mathbf{b} - \mathbf{G}\hat{\mathbf{w}}^{\textsc{mstls}}\|^2_2 / \|\mathbf{b}\|^2_2\right) + 2\log(\gamma)r(\hat{\mathbf{w}}^{\textsc{mstls}})$

        \smallskip

        \item \textbf{if} $\Delta \geq 0$: $\hat{\mathbf{w}} \leftarrow \mathbf{0}$

        \item[] \textbf{else}: $\hat{\mathbf{w}} \leftarrow \hat{\mathbf{w}}^{\textsc{mstls}}$

        \smallskip

        \item \textbf{return} $\hat{\mathbf{w}}$
    \end{enumerate}}
\end{algorithm}}

\vspace{-4mm}

It is important to note that the standard MSTLS algorithm does not natively allow for weight vectors with empty support (i.e., with $\text{supp}(\hat{\mathbf{w}}_i) = \varnothing$), and thus does not allow for trivial righthand-side models of the form \begin{align*}
    \dot{x}_i \approx \mathbf{\Theta}(\boldsymbol{x})\hat{\mathbf{w}}_i = 0.
\end{align*} To resolve this behavior in the context of the scattering ODEs of eq.~(\ref{eq:scattering_ODEs}), where indeed $\dot{\kappa}_i = 0$, we develop a `zero-admissible' extension of Algorithm~\ref{algorthim:MSTLS} inspired by the model selection work of \cite{Mangan.etal2017ProcRSocA} and \cite{Messenger.etal2024JRSocInterfacea}. In particular, we use a \textit{Bayesian information criterion} (BIC) of the form \begin{align}\label{eq:BIC}
    \mathcal{B}(\hat{\mathbf{w}}_i; \mathbf{b}_i, \mathbf{G}) := \gamma^2\log\left( \frac{\|\mathbf{b}_i - \mathbf{G}\hat{\mathbf{w}}_i\|^2_2}{\gamma^2} \right) \, + \, 2\log(\gamma)r(\hat{\mathbf{w}}_i),
\end{align} where $r(\hat{\mathbf{w}}_i) := \text{rank}(\mathbf{G}|_{\text{supp}(\hat{\mathbf{w}}_i)})$ denotes the number of uniquely estimated parameters and $\gamma^2 > 0$ is a `soft rank' heuristic for the number of statistically independent weak-form equations in eq.~(\ref{eq:weak_linear_system}), \begin{align*}
    \gamma := \frac{\|\dot{\mathbf{\Phi}}\|^2_F}{\|\dot{\mathbf{\Phi}} \dot{\mathbf{\Phi}}^T\|_F}
    \approx \sqrt{K} \frac{\|\dot{\varphi}(\mathbf{t})\|^2_2}{\|\dot{\varphi}(\mathbf{t}) \star \dot{\varphi}(\mathbf{t})\|_2},
    \quad \text{where} \quad
    \gamma^2 \sim \frac{K}{m},
    \ \ \text{as} \ \
    K \rightarrow \infty.
\end{align*} For the zero model $\hat{\mathbf{w}}_i = \mathbf{0}$ (representing conserved quantities), the corresponding BIC value is given by \begin{align*}
    \mathcal{B}(\mathbf{0}; \mathbf{b}_i, \mathbf{G}) = \gamma^2 \log\left( \frac{\|\mathbf{b}_i\|^2_2}{\gamma^2} \right),
\end{align*} meaning that \begin{align*}
    \Delta_i := \mathcal{B}(\hat{\mathbf{w}}_i; \mathbf{b}_i, \mathbf{G}) - \mathcal{B}(\mathbf{0}; \mathbf{b}_i, \mathbf{G})
    = \gamma^2\log\left(\frac{\|\mathbf{b}_i - \mathbf{G}\hat{\mathbf{w}}_i\|^2_2}{\|\mathbf{b}_i\|^2_2}\right) \, + \, 2\log(\gamma)r(\hat{\mathbf{w}}_i).
\end{align*} In principle, for $\Delta_i \geq 0$, trivial dynamics are preferred from an information-theoretic point of view; conversely, the non-zero MSTLS estimate is preferred whenever $\Delta_i < 0$. With this in mind, our final model weights $\hat{\mathbf{w}}_i$ are determined using Algorithm~\ref{algorthim:BIC_MSTLS} given above. Numerically, we guard against degenerate weak dynamics $\|\mathbf{b}_i\|_2 \approx 0$ by defaulting to the zero model whenever $\|\mathbf{b}_i\|_2 \leq \tau_{\mathbf{b}} := \texttt{1e-8}$. Otherwise, we trigger the BIC test of Algorithm~\ref{algorthim:BIC_MSTLS} whenever the MSTLS fit is poor -- in particular, when the relative residual $\|\mathbf{b}_i - \mathbf{G}\hat{\mathbf{w}}_i^{\textsc{mstls}}\|_2 / \|\mathbf{b}_i\|_2$ exceeds a default cutoff of $0.5$.


\newpage

\section{Properties of Lax Pairs}\label{appendixB:lax_pairs}
Two linear differential operators $L$ and $M$, acting on test functions $\phi = \phi(x;t)$ and potentially depending on a one-parameter family of functions $u = u(x;t)$, are called a \textit{Lax pair} if they satisfy \textit{Lax's equation}, \begin{align*}
    L_t = [M,L],
    \quad \text{or equivalently,} \quad
    L_t + [L,M] = 0,
\end{align*} just as in eq.~(\ref{eq:lax_eqn}) above. In many cases of interest, $L$ is self-adjoint while $M$ is skew-adjoint, in which case the point spectrum of $L$ is invariant in time (i.e., $L(u) \sim L(u_0)$ for every $t$). This can be explicitly shown by considering the EVP given by $L\phi = \lambda\phi$ with normalized eigenfunctions $\|\phi\|_2 = 1$, so that $\lambda = \langle \phi, L\phi\rangle$. Differentiating with respect to time then yields \begin{align*}
    \dot{\lambda}
    = \langle \phi, L_t\phi\rangle
    = \langle \phi, [M,L]\phi\rangle
    = \lambda\langle \phi, M\phi\rangle - \lambda\langle \phi, M\phi\rangle
    = 0,
\end{align*} where the second equality above follows from Lax's equation and the third uses $\langle \phi, ML\phi\rangle = \lambda\langle\phi, M\phi\rangle$ together with $\langle\phi, LM\phi\rangle = \lambda\langle\phi, M\phi\rangle$. Moreover, for any eigenpair $(\lambda, \phi)$ satisfying $L(u_0)\phi_0 = \lambda\phi_0$ at the initial time $t = 0$, the transported eigenfunction $\phi := P(t)\phi_0$ satisfies the linear system of eq.~(\ref{eq:lax_evp}), where $P(t)$ is a unitary operator representing a solution to $\dot{P} = M(u) P$ with $P_0 = I$.

\end{appendix}


\begin{Backmatter}


\paragraph{Acknowledgments}
The authors would like to thank Mark Ablowitz for his helpful insights, especially concerning future work with Painlev\'e equations. This work utilized the Blanca condo computing resource at the University of Colorado Boulder. Blanca is jointly funded by computing users and the University of Colorado Boulder.

\paragraph{Funding Statement}
This work is supported in part by National Science Foundation Grants 2054085 and 2109774, National Institute of Food and Agriculture Grant 2019-67014-29919, and Department of Energy Grant DE-SC0023346.

\paragraph{Competing Interests}
The authors declare no competing interests.

\paragraph{Data Availability Statement}
Replication data and code are publicly available: \href{https://github.com/MathBioCU/Scattering}{https://github.com/MathBioCU/Scattering}

\paragraph{Ethical Standards}
The research meets all ethical guidelines, including adherence to the legal requirements of the study country.

\paragraph{Author Contributions}
Conceptualization: S.M; V.D.; D.M.B. Methodology: S.M; V.D.; D.M.B. Data curation: S.M. Data visualization: S.M. Writing original draft: S.M. All authors approved the final submitted draft.



\bibliography{references}

\end{Backmatter}

\end{document}